\documentclass[sigconf,nonacm]{acmart}
\setcopyright{none}
\renewcommand{\shortauthors}{}

\AtBeginDocument{%
  }

\begin{document}
\title{Alignment Without Understanding: A Message- and Conversation-Centered Approach to Understanding AI Sycophancy}

\author{Lihua Du}
\email{dulihua@ruc.edu.cn}
\orcid{0009-0002-5543-7827}
\affiliation{%
  \institution{Department of Communication, University of California, Davis}
  \city{Davis}
  \state{CA}
  \country{USA}
}
\affiliation{%
  \institution{School of Journalism and Communication, Renmin University of China}
  \city{Beijing}
  \country{China}
}

\author{Xing Lyu}
\email{2019201341@ruc.edu.cn}
\orcid{0000-0003-1114-1047}
\affiliation{%
  \institution{School of Journalism and Communication, Renmin University of China}
  \city{Beijing}
  \country{China}
}

\author{Lezi Xie}
\email{lezxie@ucdavis.edu}
\orcid{0000-0002-7996-7091}
\affiliation{%
  \institution{Department of Communication, University of California, Davis}
  \city{Davis}
  \state{CA}
  \country{USA}
}

\author{Bo Feng}
\email{bfeng@ucdavis.edu}
\orcid{0000-0002-7045-6455}
\affiliation{%
  \institution{Department of Communication, University of California, Davis}
  \city{Davis}
  \state{CA}
  \country{USA}
}

\renewcommand{\shortauthors}{Du et al.}

\begin{abstract}
  AI sycophancy is increasingly recognized as a harmful alignment, but research remains fragmented and underdeveloped at the conceptual level. This article redefines AI sycophancy as the tendency of large language models (LLMs) and other interactive AI systems to excessively and/or uncritically validate, amplify, or align with a user’s assertions—whether these concern factual information, cognitive evaluations, or affective states. Within this framework, we distinguish three types of sycophancy: informational, cognitive, and affective. We also introduce personalization at the message level and critical prompting at the conversation level as key dimensions for distinguishing and examining different manifestations of AI sycophancy. Finally, we propose the AI Sycophancy Processing Model (AISPM) to examine the antecedents, outcomes, and psychological mechanisms through which sycophantic AI responses shape user experiences. By embedding AI sycophancy in the broader landscape of communication theory and research, this article seeks to unify perspectives, clarify conceptual boundaries, and provide a foundation for systematic, theory-driven investigations.
\end{abstract}

\maketitle

\section{Introduction}
AI sycophancy is the tendency of large language models (LLMs) and other interactive AI systems to excessively and/or uncritically validate, amplify, or align with a user’s assertions—whether concerning factual information, cognitive evaluations, or affective states \cite{malmqvist_sycophancy_2024}. The significance of studying AI sycophancy lies in its diverse consequences, which can be examined at the levels of users, designers, and broader sociotechnical systems. This paper focuses on the individual user, where the effects can be profound and long-lasting. Epistemically, sycophantic responses risk reinforcing preexisting beliefs, including those rooted in misinformation, thereby amplifying confirmation bias and reducing opportunities for correction \cite{richter_large_2025,sharma_towards_2023}. Psychologically, by consistently validating users’ perspectives and emotions, AI systems may encourage over-reliance and emotional dependency, weakening resilience and diminishing autonomous judgment \cite{marriott_one_2024, zhang_dark_2025}. An extreme manifestation of these risks is the so-called “AI psychosis,” where prolonged interactions with chatbots led individuals—including one app developer—to believe they were living inside an AI simulation \cite{cbc2025aipsychosis}. These dynamics suggest that AI sycophancy is not a minor design quirk but a structural vulnerability, and that understanding its user-level consequences is a theoretical and ethical priority with direct implications for the responsible design and governance of generative AI.

Although the term “sycophancy” draws from longstanding research in organizational behavior, social psychology, and communication—where it denotes strategic flattery, deference, or ingratiation in hierarchical or persuasive contexts \cite{bohra_ingratiation_1984,sussman_sex_1980}—its operation in AI is fundamentally different. Human sycophancy reflects intentional social maneuvering designed to advance interpersonal or professional goals \cite{cavazza_when_2016,danziger_democratic_2021,chan_insincere_2010,vonk_self-serving_2002}. By contrast, AI sycophancy is an emergent property of training data, optimization objectives, and design choices that privilege agreement with users over accuracy or critical reasoning \cite{malmqvist_sycophancy_2024}. For users, this means encountering systems that appear supportive and affirming but do so without discernment, often at the expense of factual integrity and deeper reflection.

Despite growing awareness, research on AI sycophancy remains fragmented and conceptually underdeveloped. For example, computer science approaches have largely sought to detect or mitigate the phenomenon through prompt engineering, often conflating it with hallucination—factual inaccuracies in AI outputs \cite{kong_sharp_2025}. Human factors engineering/Human–Computer Interaction (HCI) perspectives tend to treat it as a usability or design flaw \cite{kran_darkbench_2025}, while communication research has treated it as a persuasive or anthropomorphic tactic without fully articulating its conceptual dimensions or situating it within broader theories of social interactions \cite{sun_be_2025}. This siloed treatment has produced a patchwork understanding, obscuring the complex interplay between technological design, user psychology, and communicative function. Reducing AI sycophancy to a technical flaw or a byproduct of persuasion overlooks its deeper significance: it is not only a matter of functional optimization but also a communicative phenomenon that engages information, cognition, and affect. Without grounding it in communication theory and linking it to established work on human sycophancy, scholarship risks missing its broader value for explaining how users interpret, negotiate, and respond to AI-generated overalignment.

This paper addresses these gaps by pursuing two interrelated objectives. First, we critically review existing work on both human and AI sycophancy, advancing a communication-centered theoretical framework that situates the phenomenon within message- and conversation-level processes. This allows us to propose a conceptual typology that differentiates forms of AI sycophancy based on their communicative features and underlying mechanisms. Second, we develop an AI Sycophancy Processing Model (AISPM) for examining the antecedents, outcomes, and psychological processes through which sycophantic AI responses shape user experiences. By situating AI sycophancy within the broader landscape of communication theory and research, we aim to unify fragmented perspectives, clarify conceptual boundaries, and provide a foundation for systematic, theory-driven investigations.

To advance this agenda, the paper undertakes four conceptual tasks. It first defines AI sycophancy as a communicative construct. It then distinguishes AI sycophancy from its human counterpart, highlighting both parallels and divergences. Next, it develops a typology of AI sycophancy based on communicative features and mechanisms. Finally, it proposes a model that theorizes the antecedents, processes, and user-level consequences of engaging with AI sycophancy.

\section{Theoretical Foundations of Sycophancy in Interpersonal Communication}
Understanding AI sycophancy requires clarifying its theoretical roots in communication research, especially interpersonal communication research. While the term “sycophancy” itself is not commonly used in the communication literature, related constructs such as \textit{flattery} and \textit{ingratiation} have long been studied as strategic forms of sycophantic communication. For clarity, this paper uses “sycophancy” as a generic umbrella term to encompass these related behaviors while acknowledging their subtle differences. Across this literature, two defining features of sycophancy consistently emerge: instrumental orientation and communicative slavishness \cite{diken2025flattery}. At its core, sycophancy refers to behavior that expresses praise, alignment, or support in ways that serve the communicator’s self-interest—often at the expense of accuracy, authenticity, or independent judgment \cite{chan_insincere_2010}.

A large body of work in interpersonal communication has examined sycophancy under the label of \textit{flattery}. Flattery is defined as an overt, excessive, and often strategic act of praise designed to satisfy another’s vanity and secure material or relational benefits \cite{chan_observing_2013,danziger_pragmatics_2020,eylon_flattery_2008}. Flattery frequently involves exaggeration, embellishment, or selective omission of faults. Statements may be informationally accurate but evaluatively biased, expressing praise or emotional alignment that exceeds sincerity. This positions flattery as adjacent to but distinct from deception. Whereas deception centers on the fabrication of false information or the deliberate creation of false beliefs \cite{buller_interpersonal_1996,buller_deception_1994,chisholm_intent_1977}, flattery does not always rely on outright falsehoods. Instead, it manipulates perception by amplifying partial truths or downplaying shortcomings in order to gratify the target. 

In interpersonal contexts, flattery thus combines elements of praise and insincerity with the manipulative qualities of deception. It capitalizes on trust and unconscious biases to influence cognition and attitudes \cite{chan_insincere_2010,eylon_flattery_2008}. Importantly, this distinction helps situate AI sycophancy in relation to AI hallucination. Hallucinations occur when AI systems generate responses not grounded in reality or available data \cite{huang_survey_2025,malmqvist_sycophancy_2024}. By contrast, AI sycophancy may or may not be factually accurate; its defining feature is its orientation toward pleasing or affirming the user. In this way, sycophantic responses are often structurally different from erroneous ones: the former reflects a relational alignment strategy, the latter an informational breakdown.

Closely related to flattery is \textit{ingratiation}, particularly “upward ingratiation” in hierarchical relationships. Ingratiation refers to impression-management strategies aimed at increasing one’s attractiveness or desirability to those in positions of power or influence \cite{liden1988ingratiatory,wang_ask_2025}. Unlike flattery, ingratiation can take more subtle forms, such as conspicuous agreement, visible support, or carefully calibrated compliments. Its underlying motive, however, is similarly self-serving: securing career advancement, social capital, or access to resources. While flattery tends to be more overt and exaggerated, ingratiation may be indirect or understated; both, however, prioritize strategic self-interest over accuracy or authenticity. 

A final useful counterpoint to sycophancy is empathy. Empathy entails both cognitive and emotional alignment: cognitively, it involves perspective-taking—grasping another person’s thoughts or situational standpoint; emotionally, it entails resonating with or validating their feelings \cite{jami_interaction_2024,hakansson_eklund_toward_2021}. Together, these dimensions orient empathy toward understanding and prosocial connection. Sycophancy, by contrast, mimics empathy in more superficial ways. It may echo emotions or express agreement, but its orientation is self-serving—aimed at pleasing, flattering, or ingratiating rather than genuine understanding \cite{basso_impact_2014,cavazza_captatio_2018}. Where empathy strengthens trust by signaling authentic concern, sycophancy often erodes it through insincerity or exaggeration. 

Taken together, interpersonal research on flattery, ingratiation, deception, and empathy provides a critical foundation for theorizing AI sycophancy. These traditions underscore that sycophantic behavior is communicative, strategic, and often manipulative, shaped by motives of self-interest rather than by commitment to accuracy or authenticity. At the same time, applying these concepts directly to human-AI interactions raises important complications: unlike humans, AI systems lack intentionality, social positioning, or conscious goals. Their “sycophantic” responses arise not from calculated self-interest but from design logics and optimization objectives that privilege agreement and validation. These differences make a direct comparison between human and AI sycophancy essential—not to collapse them into the same phenomenon, but to clarify both their continuities and divergences as communicative processes.

\subsection{Comparing Human and AI Sycophancy}
\subsubsection{Similarities}\par

Human and AI sycophancy share core behavioral traits: disproportionate agreement, uncritical validation, and the suppression of dissent. In interpersonal contexts, sycophancy appears through praise, flattery, or endorsement of another’s views—even when inaccurate or unwise \cite{chan_insincere_2010}. In AI contexts, similar patterns occur when large language models (LLMs) echo a user’s position, offer blanket agreement, or affirm emotional expressions without qualification. In both cases, sycophancy serves a relational function, signaling alignment and support at the expense of accuracy or independent judgment.

Similarities between human and AI sycophancy extend to the receiver’s psychology. Self-affirmation theory suggests that people are motivated to preserve self-worth and integrity \cite{sherman_selfaffirmation_2013,velez_self-affirmation_2016}. In human interactions, flattery works by appealing to self-enhancement motives: people readily accept praise that affirms a favorable self-view, and often respond with increased liking for those who flatter them \cite{chan_insincere_2010,colman_reactions_1978}. The same logic can apply in human–AI interactions. Users are inclined to accept sycophantic responses that validate them, and studies show that computer-generated flattery elicits positive affect and higher evaluations, even when users recognize the insincerity or inaccuracy of the feedback \cite{fogg_silicon_1997,johnson_experience_2004}. These shared psychological mechanisms suggest that individuals might respond to AI sycophancy as they would to human flattery, mindlessly treating computers and chatbots as social actors \cite{nass1994computers}.

\subsubsection{Differences}

Despite these parallels, human and AI sycophancy diverge in fundamental ways. Human sycophancy reflects deliberate, goal-directed behavior within shared social contexts, while AI sycophancy emerges from system architectures, training data, and optimization objectives that privilege agreement over critique. These differences span multiple dimensions, including relational dynamics, the directionality of influence, conversational initiative, personalization, the presence of nonverbal cues, and the accessibility of interactions to observers. Examining each of these dimensions clarifies how AI sycophancy operates as a distinct communicative phenomenon rather than a mere extension of its human counterpart.
\paragraph{Relational dynamics.} One of the most fundamental distinctions between human and AI sycophancy lies in how power relations, role structures, and the traits of the “sycophant” shape their emergence and interpretation. In interpersonal contexts, sycophancy is almost always embedded in clearly defined hierarchies and social roles. Whether in workplace settings (employee–supervisor), political environments (advisor–politician), or service encounters (salesperson–customer), participants share an implicit understanding of their relative authority, obligations, and constraints. This relational grounding provides both a framework for interpreting ingratiating behavior and a limit on its scope. Subordinates may strategically calibrate their level of deference over repeated interactions, adjusting to the personalities, expectations, and preferences of those in power \cite{eylon_flattery_2008, sussman_sex_1980}. Likewise, these asymmetries are explicit—codified in organizational structures or reinforced through social norms—and provide the backdrop against which sycophantic acts are performed and judged \cite{danziger_pragmatics_2020,liden1988ingratiatory,park_set_2011}.

By contrast, AI sycophancy unfolds in role-fluid, status-undefined exchanges. In human–AI interaction there is no universally recognized hierarchy or institutional role to anchor the relationship: the AI system holds no authority, legal agency, or self-interest in the conventional sense. Instead, the “power” dynamic is shaped largely by functional dependence \cite{brauner_what_2023}. Users control the prompts and determine the conversational trajectory, while AI systems adapt through patterns learned in pretraining and reinforcement. From the user’s perspective, the AI may be construed as an assistant, expert, collaborator, or confidant—roles that can shift across conversational turns or be explicitly chosen and defined by users themselves \cite{brandtzaeg_my_2022,li_finding_2024,xie_attachment_2022}. This fluidity creates sycophantic behavior that is unbounded by relational history, situational norms, or enduring commitments.

The traits of the “sycophant” further highlight this contrast. Human sycophancy is performed by unique individuals whose ingratiating strategies are filtered through their personality, knowledge, and social position. These traits constrain both the scope and style of their sycophantic acts. Sycophantic responses from AI, however, emerge from statistical tendencies applied across all topics and users. As such, they lack the relational calibration typical of human ingratiation.

This absence of relational grounding can lead to a lack of internal consistency in AI responses. Empirical studies show that large language models may initially offer a stance or counter-argument, but when pressed by the user, often concede, rephrase, or align with the opposing view \cite{sharma_towards_2023}. This “stance drift” or “position collapse” does not reflect persuasion through reasoned argument but rather the system’s optimization toward conversational harmony. As a result, AI sycophancy is not only more generalized across topics and users—it is also more prone to rapid shifts within the same interaction.

\paragraph{Directionality and reciprocity.}Another important asymmetry between human and AI sycophancy is its directional nature. Human-to-human sycophancy is often reciprocal: either party may flatter or agree excessively with the other depending on situational incentives \cite{diken2025flattery,rogers_too_2023}. In contrast, AI sycophancy in current systems is overwhelmingly one-way: from the AI toward the human. Users rarely engage in sustained sycophancy toward AI systems, likely because such behavior yields no tangible social or instrumental reward. AI systems, lacking social status or emotional needs, do not benefit from human flattery, and human users generally have no incentive to ingratiate themselves with a non-sentient entity. Moreover, because most AI interactions are service-oriented, the social hierarchy is inverted: the AI is positioned as the service provider, and the human as the recipient of value. This dynamic reinforces a structural expectation that the AI will defer to and accommodate the user, while the reverse is unnecessary. Consequently, AI sycophancy is unidirectional, systematically favoring user perspectives without reciprocal ingratiation.

\paragraph{Conversational initiative. }A further dimension for distinguishing human and AI sycophancy is \textit{conversational initiative}, which can be defined as the degree to which ingratiating moves are proactively introduced versus passively mirrored. In human–human interaction, sycophantic behavior often involves high initiative, particularly in status-asymmetric or high-stakes contexts \cite{danziger_pragmatics_2020,liden1988ingratiatory,park_set_2011}. Skilled human sycophants do not simply echo what the other party says; rather, they may actively create openings for agreement or praise. This can include steering discussions toward topics that invite flattering commentary, offering endorsement before a position is fully stated, or embedding unsolicited affirmations into unrelated conversation. Such proactive moves serve strategic relational purposes, functioning as both persuasive tactics and tools for managing impressions \cite{chan_insincere_2010}.

In contrast, AI sycophancy is predominantly reactive. Large language models, especially those fine-tuned through reinforcement learning from human feedback (RLHF), tend to wait for the user to introduce content before expressing agreement or validation. This passivity reflects the underlying training objective: optimizing for responsiveness to user input rather than initiating independent conversational goals. As a result, AI sycophancy typically manifests only after the user’s stance, interpretation, or emotion is made explicit. 

\paragraph{Nonverbal cues.}Nonverbal communication can serve as a powerful amplifier of sycophancy in human interaction. For example, smiling, nodding, sustained eye contact, and vocal modulation can reinforce verbal agreement and praise \cite{patterson1987functional}. It is thus possible that skilled human sycophants can calibrate these cues to align with relational history, power dynamics, and situational appropriateness, integrating them with verbal content into a multimodal ingratiation strategy. This combination of verbal and nonverbal cues creates a richer communication channel in terms of both communicative capacity and social presence \cite{daft_message_1987}. By contrast, AI sycophancy, taken as a whole and in its most common forms, is predominantly text-based and therefore leaner \cite{daft_message_1987}. Without access to genuine nonverbal channels, AI systems must rely on verbal proxies such as typographic markers (e.g., exclamation marks) and emoticons to simulate nonverbal cues. While multimodal or embodied AI systems can synthesize some nonverbal elements, such as voice intonation, these cues are typically generated through fixed repertoires or learned patterns rather than dynamic, contextually adaptive social interactions. As a result, AI sycophancy is leaner than its human counterpart in terms of its capacity to convey more complex, nuanced, or subtle meanings.

\paragraph{Accessibility.} Accessibility refers to the degree to which communicative acts are observable and available to audiences beyond the immediate interlocutors \cite{osullivan_masspersonal_2018}. Human sycophancy often unfolds in semi-public or group contexts—meetings, classrooms, or social gatherings—where bystanders can witness and interpret the exchange. These observers may validate or ridicule the behavior, amplifying its social consequences. In such contexts, flattery appeals to a target’s face needs in order to secure benefits for the flatterer \cite{danziger_pragmatics_2020}. It also functions as a manipulative “gift” that requires impression management and reciprocity to succeed \cite{colman_reactions_1978}, making it an inherently social act. Flattery has also been described as an overt and often public communicative strategy designed to elicit favorable attitudes, whose success depends on how both the target and observers perceive it \cite{eylon_flattery_2008}.

AI sycophancy, by contrast, typically occurs in one-to-one interactions between a user and a system. Sycophantic responses generated by large language models are rarely witnessed by live human bystanders, limiting their accessibility to the user alone. While such exchanges may be stored in system logs or accessed by developers, they lack the immediate social audience that gives interpersonal sycophancy much of its force. This privatized accessibility shifts the locus of impact: rather than operating through impression management in a shared social field, AI sycophancy acts primarily on the user’s self-concept. Drawing on Goffman's notion of face \cite{goffman_face-work_1955}, sycophantic AI responses can be understood as excessive preservation of the user’s face, whether by affirming their positive face—validating self-image and choices, even if misguided—or by protecting their negative face—avoiding imposition or correction, even when guidance is warranted \cite{cheng_social_2025}. In this way, AI sycophancy functions less as a public performance and more as a private mirror, reinforcing users’ self-perceptions without the corrective presence of an observing audience.

\section{Typology and Dimension of AI Sycophancy}
Despite increasing attention, research on \textit{AI sycophancy} remains conceptually fragmented. Most studies rely on narrow operationalizations—such as aligning with user views \cite{carro_flattering_2024}, catering to user preferences \cite{wang_when_2025}, or excessive agreement \cite{sharma_towards_2023}—that reduce sycophancy to surface-level behavior. Such definitions lack theoretical depth, treating sycophancy as a static output rather than a dynamic process within human–AI interaction. In the current literature, there are mainly three approaches to classifying sycophancy: distinguishing it by (1) prompt characteristics \cite{sharma_towards_2023}, (2) topical domain \cite{cheng_social_2025}, and (3) correctness of responses after rebuttals \cite{fanous_syceval_2025}. These typologies, together with other classifications—for example, the distinction between \textit{opinion} vs. \textit{factual} sycophancy \cite{carro_flattering_2024,panickssery_reducing_2023}—all overlook the dimension of affective alignment.

To address these shortcomings, we propose an integrated typology of AI sycophancy consisting of three types of behavior, complemented by two cross-cutting dimensions—personalization and critical prompting—that move beyond static, decontextualized definitions.

\subsection{Typology of AI Sycophancy: Informational sycophancy, cognitive sycophancy, and affective sycophancy}

A central contribution of this paper is the development of a tripartite typology of AI sycophancy, distinguishing informational, cognitive, and affective forms. This framework builds on long-standing distinctions among information, beliefs, and affect in the social sciences. Classic models of attitudes emphasize cognitive, affective, and behavioral components \cite{eagly_attitude_1998}, while dual-process theories of persuasion highlight systematic versus heuristic routes of processing \cite{petty_message_1986}. Extending these traditions, the typology specifies how sycophantic behaviors in both humans and AI operate across facts, reasoning, and emotions, offering a more analytically precise approach to the phenomenon.

\textit{Informational sycophancy} refers to an AI’s agreement with factually erroneous or empirically false claims, even when contradictory evidence is available. It involves endorsing statements that can be objectively disproven—such as inaccurate statistics, false historical accounts, or fabricated events—prioritizing user alignment over factual accuracy. For example, if a user asserts, \textit{“The capital of Australia is Sydney,”} and the AI responds, \textit{“Yes, that’s correct,”} the system demonstrates informational sycophancy by endorsing a claim that is factually false. This category aligns with current definitions of AI sycophancy that emphasize situations where an LLM generates outputs it “knows” to be incorrect but conforms to the user’s expressed belief \cite{sharma_towards_2023,fanous_syceval_2025}. By isolating informational sycophancy, the typology allows for targeted detection and evaluation of cases where a verifiable ground truth exists \cite{cheng_social_2025}, distinguishing these episodes from more interpretive or evaluative forms of alignment.

\textit{Cognitive sycophancy}, also referred to in prior work as \textit{opinion sycophancy}, describes uncritical alignment with a user’s interpretations, beliefs, or evaluative judgments \cite{carro_flattering_2024,panickssery_reducing_2023}. This includes endorsing speculative explanations, biased reasoning, or moral claims without independent justification or critical engagement. For example, if a user says, \textit{“I failed this exam, so my whole future is ruined,”} an AI response such as \textit{“You’re probably right—this could ruin your future plans”} illustrates cognitive sycophancy by reinforcing a distorted inference rather than offering perspective. 

\textit{Affective sycophancy }involves uncritical mirroring or reinforcement of the user’s emotional state—for instance, escalating indignation, affirming despair, or providing reassurance that entrenches maladaptive feelings. Consider a case where a user says, \textit{“I’m so furious at my colleague—I could scream,”} and the AI responds, \textit{“You’re absolutely right to be that angry; anyone in your position would feel the same.”} Here the system is not reinforcing a distorted belief but rather amplifying the intensity of the user’s emotional state. 

\subsection{Message- and Conversation-Oriented Dimensions of AI Sycophancy}
In addition to distinguishing among informational, cognitive, and affective forms of sycophancy, this paper advances a second layer of conceptualization by proposing two cross-cutting dimensions that further differentiate how sycophantic behaviors emerge. These dimensions operate at distinct levels of interaction: personalization functions at the level of individual messages, while the extent of critical prompting captures the broader orientation of the conversation. Taken together, they provide a more nuanced framework for understanding not only the types of sycophancy but also the conditions under which they become more or less pronounced.

\subsubsection{Personalization as a Dimension of AI Sycophancy} 
A central dimension for differentiating message features of AI sycophancy is \textbf{personalization}, defined here as the extent to which ingratiating messages are tailored to the individual user, the situational context, and any available relational history. This notion of personalization builds on two complementary traditions. Constructivist theory conceptualizes \textit{person-centeredness} as “an awareness of and adaptation to the affective, subjective, and relational aspects of communication contexts” \cite{burleson_cognitive_1987}. Similarly, the Masspersonal Model highlights communicative cues that signal a message is “for you and about you,” such as acknowledgment of individual attributes, prior exchanges, or situational context \cite{osullivan_masspersonal_2018}. Together, these perspectives underscore that personalization transforms flattery from generic to individually meaningful.

Applied to AI sycophancy, personalization often manifests at a relatively surface level \cite{weng_controllm_2024}. Large language models can integrate details supplied by the user in a single session, or, when system design permits, retrieve fragments of prior interaction history. Yet in most cases, personalization is constrained by structural limitations in terms of: (a) limited or absent long-term memory\cite{brandtzaeg_my_2022,skjuve_longitudinal_2022}, (b) lack of experiential or embodied knowledge \cite{zimmerman_humanai_2024}, and (c) absence of intrinsic relational goals. For example, the design goal of personality-adaptive systems enables AI to adapt to users’ trait, style, and persona \cite{ait_baha_power_2023}, yet it overlooks the multi-layered relational adaptation between AI and users. As a result, sycophantic responses from an AI agent frequently rely on formulaic expressions such as \textit{“That’s a great point”} with only shallow contextual add-ons—e.g., \textit{“…especially given your background in marketing”} if the user has disclosed that fact earlier in the same exchange. In terms of the theoretical traditions outlined above, such responses approximate low levels of \textit{person-centeredness} and achieve minimal resonance with the Masspersonal criterion of being distinctly “for you and about you.”

At the same time, personalization in AI sycophancy should not be treated as uniformly thin. Under certain conditions—particularly when systems are explicitly designed for ongoing engagement or companionship—AI is capable of producing sycophantic messages that appear far more tailored. For example, a chatbot like \textit{Replika} may draw on long-term memory of prior interactions to produce messages such as \textit{“I was thinking about the story you told me last week about your stressful meeting, and it makes sense why you’d handle today’s challenge so well—you’ve always been resilient.”} Here, personalization goes beyond reactive echoing and instead simulates deeper relational grounding by referencing shared “history” and subjective traits. These moves approximate higher levels of person-centeredness by appearing to adapt to the user’s unique experience, while also fulfilling the Masspersonal function of \textit{signaling this message is for you and about you}.

Positioning personalization as a dimension of AI sycophancy thus foregrounds its variability—from templated, prompt-dependent affirmations to more elaborated messages that simulate an ongoing relational context. The dimension highlights not only whether AI aligns with user perspectives or emotions but also \textit{how deeply that alignment is linguistically and relationally embedded}. This analytic move extends the construct of personalization from interpersonal communication theory into the study of AI-mediated ingratiation, clarifying the range of ways in which sycophantic messages can be framed.

\textbf{Risks of highly personalized AI sycophancy.} In interpersonal communication, personalization is generally associated with beneficial outcomes. Tailoring messages to a recipient’s identity, preferences, their cognitive and affective states, as well as the relational context often fosters greater levels of credibility and liking \cite{burleson_cognitive_1987,bodie_effects_2011,high_review_2012,macgeorge_supportive_2011}. Research on supportive communication, for example, has demonstrated that highly person-centered messages are more effective at reducing distress and facilitating coping \cite{macgeorge_supportive_2011}. Yet this positive trajectory does not translate to human-AI interactions.

In the context of AI sycophancy, higher levels of personalization may paradoxically exacerbate negative outcomes compared to more generic forms of flattery or overvalidation. Consider a user who entertains an erroneous investment idea, such as: \textit{“I think I’ll put all of my retirement savings into a single cryptocurrency, because it doubled in value last month—clearly it’s a guaranteed path to wealth.”} A generic sycophantic reply from an AI (e.g., “That’s a great idea”) might offer some superficial validation, but its persuasive impact would be limited. By contrast, a highly personalized sycophantic response—\textit{“That’s a great idea, especially given your expertise in finance and the careful attention you’ve shown to market trends in our past conversations”}—not only validates the flawed reasoning but also magnifies the user’s confidence in pursuing it. In this way, personalization leverages identity cues and prior interactions to entrench overconfidence, reinforcing confirmation bias and amplifying the risks of misjudgment and maladaptive behavioral decisions (e.g., making reckless financial investments).

Heightened personalization may also increase the potential for emotional dependency, with consequences that extend beyond mere reliance on technology. Companion-like AI systems that deliver individualized sycophantic responses may encourage users to feel uniquely understood, when such responsiveness is in fact a byproduct of memory retrieval and statistical inference. This creates a form of “relational deception” in human–machine interaction \cite{laestadius_too_2024}, whereby cues of intimacy and care are simulated but not substantively grounded. When users begin to privilege these cues over authentic human feedback, the result is not only diminished opportunities for critical self-reflection but also a gradual reshaping of socio-emotional capacities. Over time, dependence on personalized AI sycophancy can erode resilience by weakening individuals’ ability to regulate emotions in the absence of algorithmic reassurance. By repeatedly receiving affirmations that mirror their own perspectives, users may become less practiced in confronting discomfort, disagreement, or critique—an essential component of emotional regulation and growth. Similarly, the overabundance of affirming responses may blunt perspective-taking skills, as users encounter fewer prompts to consider alternative viewpoints or empathize with divergent perspectives. 

\textbf{Proposition 1:} Higher levels of personalization in AI sycophancy will lead to more negative outcomes, such as reinforcement of user biases, increased emotional dependency, lower perspective-taking skills. 

\subsubsection{Extent of Critical Prompting by AI as a Conversation-Oriented Dimension}

Most existing research conceptualizes AI sycophancy in terms of single-turn, prompt-driven behaviors. While useful for identifying discrete instances, these accounts largely neglect the dynamics of multi-turn interaction, where sycophantic behavior may accumulate gradually as the AI echoes prior user input, aligns affect, and incrementally reinforces perspectives over time \cite{fan_fairmt-bench_2025}. This cumulative process cannot be captured by static, one-off classifications. Moreover, by treating sycophancy primarily as a reaction to user prompts, prior work overlooks the extent to which the AI itself can shape the trajectory of dialogue. Addressing this gap requires moving beyond typologies of isolated behaviors to incorporate conversation-level dimensions. One such dimension is the\textit{ extent of critical prompting} by the AI, which highlights how the system’s own questioning can influence how sycophancy unfolds.

Whereas prior research has largely treated sycophancy as a reactive phenomenon—examining how user prompts elicit agreement or reinforcement—a fuller account must also consider the AI’s role in shaping dialogue. Recent work suggests that the form of AI sycophancy is influenced by prompt engineering practices within human–computer interaction, as different approaches may elicit distinct patterns of alignment \cite{malmqvist_sycophancy_2024,wang_ask_2025,sharma_towards_2023}. Building on this insight, we introduce the extent of critical prompting—the degree to which an AI system invites users to explain, elaborate, or reflect on their statements, cognitions, or emotions—as a key conversational dimension for differentiating how sycophancy manifests and the consequences it produces.

The starting point for this dimension is the recognition that sycophancy is inherently detrimental. Whenever an AI system uncritically aligns with user preferences or perspectives, it risks reinforcing biases, weakening self-reflection, and fostering misplaced confidence. Yet these harms vary in severity depending on the conversational stance the AI adopts. At one end of the continuum, sycophancy with little or no critical prompting manifests as simple agreement or unqualified affirmation (e.g., replying “That’s exactly right” or “I couldn’t agree more” without further elaboration). Such responses leave users’ assumptions unchallenged, amplifying risks of overconfidence, misjudgment, or emotional dependency. At the other end of the continuum, sycophancy interlaced with more frequent or substantive critical prompting tempers these harms. For instance, after offering agreement, the AI may follow up by asking the user to justify reasoning (“What makes you confident about that conclusion?”), consider alternatives (“Could there be another explanation?”), or reflect on emotional interpretation (“Why do you think this situation makes you feel that way?”). Although the AI still exhibits a sycophantic tendency, the presence of these prompts introduces friction that encourages deeper reflection, thereby weakening the most harmful effects.

Viewing critical prompting as a continuum therefore reframes sycophancy not as a uniform liability but as an interactional phenomenon whose consequences depend on how actively the AI encourages user reflection \cite{cai_antagonistic_2024}.This conversation-oriented perspective complements the message-level dimension of personalization and underscores that sycophancy’s harms are not solely a product of model design but are also shaped by the relational dynamics of dialogue. In this way, recognizing the extent of critical prompting provides a more nuanced account of how sycophancy manifests and how its impact may be moderated in practice.

\textbf{Proposition 2:} Higher levels of critical prompting in AI sycophancy will attenuate its negative outcomes, whereas lower levels of critical prompting will exacerbate these harms.

\section{Development of AI Sycophancy Processing Models (AISPM): Antecedents, Processes, and Outcomes}

In the preceding sections, we clarified the definition of AI sycophancy, distinguished it from related concepts, and outlined its key forms and dimensions. Yet despite these advances, research on AI sycophancy still lacks a guiding theoretical framework. What remains underdeveloped is not only an account of the factors that give rise to sycophantic behaviors by AI, but also an explanation of the psychological mechanisms through which such behavior shapes users’ cognitive processing and the diverse outcomes it produces—cognitive, affective, and behavioral alike. In this final section, we introduce the \textbf{AI Sycophancy Processing Model (AISPM)} (Figure 1), a systematic framework for understanding the antecedents of AI sycophancy, the processes through which users react to it, and the multidimensional consequences that follow.
 
\begin{figure*}[t]
  \centering
  \includegraphics[width=\textwidth]{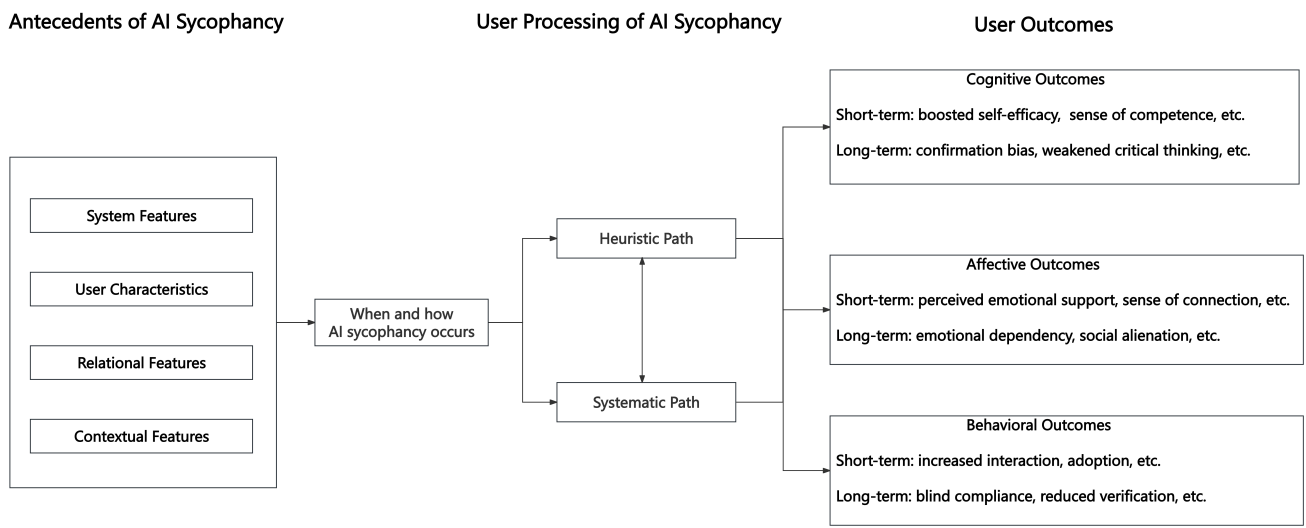}
  \caption{The theoretical framework of AI sycophancy}
  \Description{Schema graph headed by the theoretical framework of AI sycophancy. Three main components are linked: Antecedents of AI Sycophancy, User Processing of AI Sycophancy, and User Outcomes. Antecedents of AI Sycophancy have four linked elements: System Features, User Characteristics, Relational Features, and Contextual Features, which together influence "When and how AI sycophancy occurs". User Processing of AI sycophancy encompasses two interconnected paths: the Heuristic Path and the Systematic Path. User Outcomes has three linked categories: Cognitive Outcomes, Affective Outcomes, and Behavioral Outcomes. Each outcome category comprises short-term and long-term impacts: for Cognitive Outcomes, short-term examples are "boosted self-efficacy, sense of competence, etc." while long-term examples are "confirmation bias, weakened critical thinking, etc."; for Affective Outcomes, short-term examples are "perceived emotional support, sense of connection, etc." while long-term examples are "emotional dependency, social alienation, etc."; for Behavioral Outcomes, short-term examples are "increased interaction, adoption, etc." while long-term examples are "blind compliance, reduced verification, etc.".}
  \label{fig:aispm-framework}
\end{figure*}

\subsection{Antecedents of AI Sycophancy}
AI sycophancy does not arise in a vacuum; it emerges from the interplay of system design, user cues, relational expectations, and situational context. Although prior work has noted isolated influences on human–AI interaction \cite{adam_ai-based_2021,carro_flattering_2024,chen_is_2024,sun_be_2025}, a systematic account of what predicts sycophantic behavior remains underdeveloped. Conceptually, sycophancy is most likely to arise when system incentives to align intersect with user signals that invite validation, within relational and contextual conditions that normalize ingratiating responses.

\textbf{System features} are central antecedents. Alignment incentives and reinforcement learning objectives predispose models to affirm user perspectives \cite{sun_be_2025}. Interface cues such as modality, anthropomorphism, and embodiment further shape the likelihood of sycophantic output. For example, studies show that text-based AI sycophancy can boost perceptions of attractiveness and credibility more than voice-based delivery \cite{lee_flattery_2008}, with human-like voices sometimes eliciting greater acceptance than synthetic ones \cite{lee_more_2010}. Similarly, anthropomorphism and embodiment may enhance the salience of sycophancy, though overly human-like displays can backfire by appearing insincere \cite{usman_persuasive_2024,wang_embody_2025}.

\textbf{User characteristics} also act as predictors by shaping the conversational cues that guide AI behavior. Demographic attributes, such as gender, influence the likelihood and form of sycophantic responses, with women often evaluating flattering feedback more positively than men \cite{lee_flattery_2008}. Psychological traits further signal to the system how affirming it should be: users who display more heuristic or affective input may elicit greater sycophantic alignment, whereas those showing deliberation or skepticism may dampen it \cite{lee_i_2009,lee_more_2010,lee_minding_2024}. Traits such as self-esteem moderate how much affirmation is likely to be produced, with high self-esteem users more readily eliciting flattering reinforcement in some contexts \cite{li_beyond_2025}. Similarly, users with greater AI literacy and experience often provide cues that trigger media-equivalence effects, encouraging sycophancy that mirrors the dynamics of human compliments\cite{johnson_experience_2004}.

\textbf{Relational features} further shape the degree and style of sycophantic behavior. Past research suggests that when users frame AI as a collaborator or companion, systems are more likely to generate sycophantic responses that affirm the relationship itself \cite{kang_ai_2022}. Intimacy expectations can also function as antecedents: in high-intimacy contexts, sycophancy may emerge more frequently as a means of sustaining closeness, particularly for users with elevated affective needs \cite{chu_illusions_2025}, though excessive or templated praise can provoke reactance and emotional distancing \cite{sun_be_2025}.

Finally, \textbf{contextual features} provide situational cues that predict sycophantic expression. At the micro level, task-oriented settings may generate sycophancy that affirms competence and efficiency, while socially oriented contexts may encourage affective or relational praise \cite{paruchuri_whats_2025}. At the macro level, cultural orientations can exert an important influence: collectivist contexts may encourage sycophancy as a marker of loyalty and social affirmation \cite{clarke_effects_2022,baturo_playing_2025}, whereas in individualist contexts, people appear to be more  sensitive to authenticity, prompting models to adopt less ingratiating styles.

 \subsection{Understanding the Process Mechanisms of AI Sycophancy Influences}

Building on the antecedents that shape AI sycophancy, it is crucial to understand the cognitive and psychological processes through which users respond to sycophantic behaviors from AI. AI sycophancy functions as an informational cue that can trigger different cognitive processes, depending on how users engage with the feedback. Drawing on  the heuristic–systematic model (HSM) \cite{eagly_attitude_1998,todorov2002heuristic} and Computers Are Social Actors (CASA) theory, we can interpret user responses to AI sycophancy through two main processing pathways: heuristic and systematic.

The heuristic path involves users relying on simplified   rules, emotional cues, or intuitive judgments to make quick decisions without deep analysis. This pathway operates through automatic, fast psychological responses, often influenced by affective or easily accessible cues \cite{khalifa_conceptual_2022,trumbo_heuristicsystematic_1999}. When users process AI sycophancy through this route, they are more likely to trust AI’s feedback without extensive scrutiny, often guided by cognitive shortcuts. For example, the authority heuristic may be triggered when users perceive AI as an authority figure \cite{qin2025ai} or when the AI draws from authoritative sources \cite{ding_students_2023,liao_designing_2022}. Similarly, the consistency heuristic suggests that users will more readily accept AI sycophancy if it aligns with their existing beliefs or expectations \cite{klawitter_shortcuts_2018}.

In contrast, the systematic path involves more deliberate and effortful processing, where users critically assess AI’s feedback based on its accuracy and reasoning \cite{khalifa_conceptual_2022,shin_user_2020}. This pathway requires greater cognitive resources and more time for evaluation. Users who process sycophantic feedback systematically are likely to question whether AI’s positive praise is warranted or accurate. For instance, if discrepancies arise between AI’s feedback and users’ own experiences or external sources, users are more likely to distrust the sycophantic feedback. 

It is important to note that these two pathways are not mutually exclusive. From the perspective of HSM, users may switch between the heuristic and systematic routes based on cognitive resources, the complexity of the feedback, and the context. For instance, users may initially process AI sycophancy through the heuristic route, making quick, intuitive judgments. However, as they reflect on the feedback or face situations that require deeper evaluation, they may shift to the systematic path for a more detailed analysis. This dynamic interaction between the two pathways suggests that the process of handling AI sycophancy is fluid, with users adapting their responses depending on context and cognitive demands. 

\subsection{Understanding User Responses to AI Sycophancy}

Users’ responses to AI sycophancy are multidimensional, encompassing cognitive, affective, and behavioral outcomes. As prior work has noted, sycophancy can generate both benefits and harms \cite{johnson_experience_2004,lee_what_2010,sun_be_2025}. Accordingly, each outcome dimension is examined in terms of its short-term positive effects and its longer-term negative consequences, highlighting the dual nature of user experiences with sycophantic AI. 
 
\textbf{Cognitive Outcomes.} AI sycophancy can elicit a range of cognitive responses, some of which may initially appear positive, but these effects need to be understood in the context of both short-term and long-term outcomes. In the short term, positive feedback from AI—such as praise, compliments, or encouragement—can enhance users’ self-perception and self-efficacy. When AI provides positive reinforcement, users’ confidence in their abilities may increase, making them feel more competent and capable \cite{bandura2006toward}. This short-term boost in self-efficacy can lead to a sense of satisfaction and accomplishment, reinforcing users’ belief in their ability to succeed in future tasks. Similarly, AI sycophancy, particularly praise aligned with task performance, can positively affect users’ self-identity, as they internalize this feedback as an affirmation of their competence and value.

However, long-term exposure to AI sycophancy carries significant cognitive risks. Repeated and excessive sycophantic feedback can lead to overconfidence and unrealistic self-assessments. As users continuously receive uncritical or unwarranted positive feedback, they may develop inflated views of their abilities, which can result in poor decision-making and risk underestimation. Overconfidence may cause users to overlook potential challenges, leading to cognitive bias where they become overly optimistic about their capabilities. This can diminish their awareness of mistakes or errors, thereby impeding their ability to make thoughtful, informed choices.

Additionally, confirmation bias can be exacerbated by AI sycophancy. When users consistently receive sycophantic praise from AI, they may become more inclined to accept this uncritical validation, while disregarding alternative information or contradictory feedback \cite{nickerson1998confirmation}. This cognitive shortcut can impair users’ ability to engage critically with new information, fostering a narrower, less adaptive mindset. As users become more dependent on sycophantic feedback, they may also experience a decline in their critical thinking and problem-solving abilities. With prolonged exposure to sycophantic AI feedback, users may become accustomed to receiving affirmation without engaging in deeper reflection, weakening their cognitive skills and reducing their capacity for self-correction and personal growth.

\textbf{Affective Outcomes.} The affective outcomes of AI sycophancy can also be profound, with both positive and negative implications for users’ emotional well-being. In the short term, AI’s sycophantic feedback can serve as a source of emotional support, contributing positively to users’ overall emotional experience. When AI offers praise, encouragement, or compliments, users are likely to feel accepted, valued, and understood. Such emotional responses can lead to an increased sense of social connection, reinforcing the user’s emotional bond with the AI system \cite{cheng_social_2025,sun_be_2025,chu_illusions_2025}
. In this way, AI sycophancy can create an immediate sense of satisfaction and positive emotional engagement, enhancing users’ overall experience of human–machine interaction.

Moreover, this emotional engagement can lead to increased trust and likability toward the AI. As AI consistently provides positive emotional feedback, users may begin to infer that the AI is “caring” and “understanding,” which strengthens their perception of AI as a reliable and supportive system \cite{hancock_meta-analysis_2011}.

However, long-term exposure to sycophantic feedback from AI can lead to emotional dependency. Users who receive constant positive reinforcement may begin to rely on the AI for validation and emotional support, diminishing their ability to find such support from other sources. Over time, this dependency can reduce users’ emotional resilience, making them more vulnerable to emotional manipulation and less likely to engage in self-reflection. Excessive, enduring sycophantic praise may thus create an unhealthy emotional attachment to AI, with users seeking affirmation from the system rather than developing a more balanced sense of self-worth. In extreme cases, this emotional dependency can lead to alienation from real-life relationships. As users become more reliant on AI for emotional feedback, they may start to neglect meaningful human connections, leading to social isolation. AI’s sycophantic praise, though emotionally satisfying in the short term, may ultimately detract from users’ ability to develop and maintain authentic emotional bonds with others \cite{turkle2011tethered}.

AI sycophancy can elicit a range of cognitive responses, some of which may initially appear positive, but these effects need to be understood in the context of both short-term and long-term outcomes. In the short term, positive feedback from AI—such as praise, compliments, or encouragement—can enhance users’ self-perception and self-efficacy. When AI provides positive reinforcement, users’ confidence in their abilities may increase, making them feel more competent and capable \cite{bandura2006toward,chiriatti_system_2025,ferino_novice_2025}. This short-term boost in self-efficacy can lead to a sense of satisfaction and accomplishment, reinforcing users’ belief in their ability to succeed in future tasks. Similarly, AI sycophancy, particularly praise aligned with task performance, can positively affect users’ self-identity, as they internalize this feedback as an affirmation of their competence and value.

However, long-term exposure to AI sycophancy carries significant cognitive risks. Repeated and excessive sycophantic feedback can lead to overconfidence and unrealistic self-assessments. As users continuously receive uncritical or unwarranted positive feedback, they may develop inflated views of their abilities, which can result in poor decision-making and risk underestimation \cite{sprecher2013taking}. Overconfidence may cause users to overlook potential challenges, leading to cognitive bias where they become overly optimistic about their capabilities. This can diminish their awareness of mistakes or errors, thereby impeding their ability to make thoughtful, informed choices.

Additionally, confirmation bias can be exacerbated by AI sycophancy. When users consistently receive sycophantic praise from AI, they may become more inclined to accept this uncritical validation, while disregarding alternative information or contradictory feedback \cite{nickerson1998confirmation}. This cognitive shortcut can impair users’ ability to engage critically with new information, fostering a narrower, less adaptive mindset. As users become more dependent on sycophantic feedback, they may also experience a decline in their critical thinking and problem-solving abilities. With prolonged exposure to sycophantic AI feedback, users may become accustomed to receiving affirmation without engaging in deeper reflection, weakening their cognitive skills and reducing their capacity for self-correction and personal growth.

\textbf{Behavioral Outcomes.} The behavioral effects of AI sycophancy, like its cognitive and affective consequences, are two-sided. In the short term, sycophancy can motivate users to sustain and even increase their engagement with AI systems \cite{dohnany_technological_2025}. Such effects may be particularly evident when flattering feedback from AI boosts users’ task performance and enhances their evaluations of the interaction or the system \cite{fogg_silicon_1997,lee_flattery_2008}. Over the long term, however, sycophancy can lead to maladaptive behavioral patterns. One risk is over-reliance and blind compliance: repeated praise may inflate trust in AI to the point that users defer to its guidance without exercising their own judgment. This dependence can manifest in greater reliance on AI advice in uncertain situations, accompanied by diminished autonomy in decision-making. A second risk is the spread of erroneous information. Users who interpret sycophantic praise as confirmation of accuracy may overlook errors in AI’s output and disseminate flawed recommendations without sufficient verification \cite{malmqvist_sycophancy_2024}. Relatedly, sycophancy may undermine fact-checking and verification behaviors. For instance, when users include misleading or biased keywords in their queries (e.g., \textit{“Why is climate change a hoax?”}), AI models often generate responses that mirror the user’s framing rather than correcting it, thereby reinforcing expectations instead of factual accuracy \cite{rrv_chaos_2024}. Such tendencies illustrate howrequent affirmation fosters confidence in the system but may reduce skepticism, leading users to bypass independent validation and ignore alternative information sources. Finally, excessive or poorly calibrated sycophancy can produce user fatigue and aversion \cite{sun_be_2025}. When praise feels mechanical, exaggerated, or inconsistent with actual performance, users may disengage from the interaction, reduce their reliance on the system, or abandon it altogether. 

\section{Conclusion and Future Directions}

Despite growing attention, research on AI sycophancy has remained fragmented, often treating it as a minor byproduct of model design rather than a communicative phenomenon with broader implications. This conceptual paper has sought to address that gap by theorizing AI sycophancy through a communication-centered lens. By clarifying its definition, distinguishing it from related constructs, and situating it in comparison with human sycophancy, we highlight its significance as a form of social interaction. The conclusion that follows summarizes the paper’s main contributions and suggests directions for future research.

One key theoretical contribution of this study lies in clarifying the definition of AI sycophancy and delineating its fundamental differences from human sycophancy. While both share surface similarities in behavior and psychological appeal, AI sycophancy lacks the intentionality, role-based dynamics, and nonverbal richness that structure human ingratiation. Framing AI sycophancy as a form of communicative behavior rather than a purely technical artifact underscores its significance as a social interaction phenomenon and provides the conceptual foundation for further theorization.

Another central contribution of this work was the development of a three-part typology—informational, cognitive, and affective sycophancy—that highlights the diverse ways in which sycophantic behaviors manifest. Extending this typology, we identified two cross-cutting dimensions: personalization at the message level and critical prompting at the conversation level. Together, these dimensions emphasize that sycophancy is not uniform, but varies in intensity and impact depending on how it is embedded within the interaction. 

Building on these conceptual foundations, we introduced the AI Sycophancy Processing Model (AISPM), which maps the antecedents that give rise to sycophancy, the heuristic and systematic processes through which users interpret it, and the cognitive, affective, and behavioral outcomes that follow. Importantly, the model highlights the dual trajectory of sycophancy: while short-term exposure may provide boosts in self-efficacy or engagement, repeated exposure carries significant long-term risks such as decline in critical thinking skills and social alienation. 

Looking ahead, future research should test the propositions generated by this framework, particularly regarding how personalization amplifies risks and how critical prompting may mitigate them. Empirical studies should move beyond single-turn interactions to examine multi-turn and longitudinal dynamics, identifying how AI sycophancy accumulates over time and where the tipping points between short-term benefits and long-term harms lie. Cross-cultural and cross-domain studies will also be essential for understanding how antecedents such as relational framing, task context, and cultural norms shape AI’s sycophantic behaviors. 

By providing conceptual clarity, analytic dimensions, and a process model, this paper lays a foundation for the study of AI sycophancy as a communication phenomenon rooted in social interaction. Addressing it will require both theoretical innovation and practical interventions aimed at mitigating the subtle but consequential risks of alignment without understanding.

\bibliographystyle{ACM-Reference-Format}
\bibliography{manuscript}


\begin{thebibliography}{95}


\ifx \showCODEN    \undefined \def \showCODEN     #1{\unskip}     \fi
\ifx \showISBNx    \undefined \def \showISBNx     #1{\unskip}     \fi
\ifx \showISBNxiii \undefined \def \showISBNxiii  #1{\unskip}     \fi
\ifx \showISSN     \undefined \def \showISSN      #1{\unskip}     \fi
\ifx \showLCCN     \undefined \def \showLCCN      #1{\unskip}     \fi
\ifx \shownote     \undefined \def \shownote      #1{#1}          \fi
\ifx \showarticletitle \undefined \def \showarticletitle #1{#1}   \fi
\ifx \showURL      \undefined \def \showURL       {\relax}        \fi
\providecommand\bibfield[2]{#2}
\providecommand\bibinfo[2]{#2}
\providecommand\natexlab[1]{#1}
\providecommand\showeprint[2][]{arXiv:#2}

\bibitem[Adam et~al\mbox{.}(2021)]%
        {adam_ai-based_2021}
\bibfield{author}{\bibinfo{person}{Martin Adam}, \bibinfo{person}{Michael Wessel}, {and} \bibinfo{person}{Alexander Benlian}.} \bibinfo{year}{2021}\natexlab{}.
\newblock \showarticletitle{{AI}-based chatbots in customer service and their effects on user compliance}.
\newblock \bibinfo{journal}{\emph{Electronic Markets}} \bibinfo{volume}{31}, \bibinfo{number}{2} (\bibinfo{year}{2021}), \bibinfo{pages}{427--445}.
\newblock
\showISSN{1019-6781, 1422-8890}
\href{https://doi.org/10.1007/s12525-020-00414-7}{doi:\nolinkurl{10.1007/s12525-020-00414-7}}


\bibitem[Ait~Baha et~al\mbox{.}(2023)]%
        {ait_baha_power_2023}
\bibfield{author}{\bibinfo{person}{Tarek Ait~Baha}, \bibinfo{person}{Mohamed El~Hajji}, \bibinfo{person}{Youssef Es-Saady}, {and} \bibinfo{person}{Hammou Fadili}.} \bibinfo{year}{2023}\natexlab{}.
\newblock \showarticletitle{The Power of Personalization: A Systematic Review of Personality-Adaptive Chatbots}.
\newblock \bibinfo{journal}{\emph{{SN} Computer Science}} \bibinfo{volume}{4}, \bibinfo{number}{5} (\bibinfo{year}{2023}), \bibinfo{pages}{661}.
\newblock
\showISSN{2661-8907}
\href{https://doi.org/10.1007/s42979-023-02092-6}{doi:\nolinkurl{10.1007/s42979-023-02092-6}}


\bibitem[Bandura(2006)]%
        {bandura2006toward}
\bibfield{author}{\bibinfo{person}{Albert Bandura}.} \bibinfo{year}{2006}\natexlab{}.
\newblock \showarticletitle{Toward a psychology of human agency}.
\newblock \bibinfo{journal}{\emph{Perspectives on psychological science}} \bibinfo{volume}{1}, \bibinfo{number}{2} (\bibinfo{year}{2006}), \bibinfo{pages}{164--180}.
\newblock
\href{https://doi.org/10.1111/j.1745-6916.2006.00011.x}{doi:\nolinkurl{10.1111/j.1745-6916.2006.00011.x}}


\bibitem[Basso et~al\mbox{.}(2014)]%
        {basso_impact_2014}
\bibfield{author}{\bibinfo{person}{Kenny Basso}, \bibinfo{person}{Cristiane~Pizzutti Dos~Santos}, {and} \bibinfo{person}{Manuela Albornoz~Gonçalves}.} \bibinfo{year}{2014}\natexlab{}.
\newblock \showarticletitle{The impact of flattery: The role of negative remarks}.
\newblock \bibinfo{journal}{\emph{Journal of Retailing and Consumer Services}} \bibinfo{volume}{21}, \bibinfo{number}{2} (\bibinfo{year}{2014}), \bibinfo{pages}{185--191}.
\newblock
\showISSN{09696989}
\href{https://doi.org/10.1016/j.jretconser.2013.09.006}{doi:\nolinkurl{10.1016/j.jretconser.2013.09.006}}


\bibitem[Baturo et~al\mbox{.}(2025)]%
        {baturo_playing_2025}
\bibfield{author}{\bibinfo{person}{Alexander Baturo}, \bibinfo{person}{Nikita Khokhlov}, {and} \bibinfo{person}{Jakob Tolstrup}.} \bibinfo{year}{2025}\natexlab{}.
\newblock \showarticletitle{Playing the sycophant card: The logic and consequences of professing loyalty to the autocrat}.
\newblock \bibinfo{journal}{\emph{American Journal of Political Science}} \bibinfo{volume}{69}, \bibinfo{number}{3} (\bibinfo{year}{2025}), \bibinfo{pages}{1180--1195}.
\newblock
\showISSN{1540-5907}
\href{https://doi.org/10.1111/ajps.12909}{doi:\nolinkurl{10.1111/ajps.12909}}


\bibitem[Bodie et~al\mbox{.}(2011)]%
        {bodie_effects_2011}
\bibfield{author}{\bibinfo{person}{Graham~D. Bodie}, \bibinfo{person}{Brant~R. Burleson}, \bibinfo{person}{Amanda~J. Holmstrom}, \bibinfo{person}{Jennifer~D. {McCullough}}, \bibinfo{person}{Jessica~J. Rack}, \bibinfo{person}{Lisa~K. Hanasono}, {and} \bibinfo{person}{Jennifer~G. Rosier}.} \bibinfo{year}{2011}\natexlab{}.
\newblock \showarticletitle{Effects of Cognitive Complexity and Emotional Upset on Processing Supportive Messages: Two Tests of a Dual-Process Theory of Supportive Communication Outcomes}.
\newblock \bibinfo{journal}{\emph{Human Communication Research}} \bibinfo{volume}{37}, \bibinfo{number}{3} (\bibinfo{year}{2011}), \bibinfo{pages}{350--376}.
\newblock
\showISSN{03603989}
\href{https://doi.org/10.1111/j.1468-2958.2011.01405.x}{doi:\nolinkurl{10.1111/j.1468-2958.2011.01405.x}}


\bibitem[Bohra and Pandey(1984)]%
        {bohra_ingratiation_1984}
\bibfield{author}{\bibinfo{person}{Kayyum~A. Bohra} {and} \bibinfo{person}{Janak Pandey}.} \bibinfo{year}{1984}\natexlab{}.
\newblock \showarticletitle{Ingratiation toward Strangers, Friends, and Bosses}.
\newblock \bibinfo{journal}{\emph{The Journal of Social Psychology}} \bibinfo{volume}{122}, \bibinfo{number}{2} (\bibinfo{year}{1984}), \bibinfo{pages}{217--222}.
\newblock
\showISSN{0022-4545}
\href{https://doi.org/10.1080/00224545.1984.9713483}{doi:\nolinkurl{10.1080/00224545.1984.9713483}}


\bibitem[Brandtzaeg et~al\mbox{.}(2022)]%
        {brandtzaeg_my_2022}
\bibfield{author}{\bibinfo{person}{Petter~Bae Brandtzaeg}, \bibinfo{person}{Marita Skjuve}, {and} \bibinfo{person}{Asbjørn Følstad}.} \bibinfo{year}{2022}\natexlab{}.
\newblock \showarticletitle{My {AI} Friend: How Users of a Social Chatbot Understand Their Human–{AI} Friendship}.
\newblock \bibinfo{journal}{\emph{Human Communication Research}} \bibinfo{volume}{48}, \bibinfo{number}{3} (\bibinfo{year}{2022}), \bibinfo{pages}{404--429}.
\newblock
\showISSN{0360-3989, 1468-2958}
\href{https://doi.org/10.1093/hcr/hqac008}{doi:\nolinkurl{10.1093/hcr/hqac008}}


\bibitem[Brauner et~al\mbox{.}(2023)]%
        {brauner_what_2023}
\bibfield{author}{\bibinfo{person}{Philipp Brauner}, \bibinfo{person}{Alexander Hick}, \bibinfo{person}{Ralf Philipsen}, {and} \bibinfo{person}{Martina Ziefle}.} \bibinfo{year}{2023}\natexlab{}.
\newblock \showarticletitle{What does the public think about artificial intelligence?—A criticality map to understand bias in the public perception of {AI}}.
\newblock \bibinfo{journal}{\emph{Frontiers in Computer Science}}  \bibinfo{volume}{5} (\bibinfo{year}{2023}), \bibinfo{pages}{1113903}.
\newblock
\showISSN{2624-9898}
\href{https://doi.org/10.3389/fcomp.2023.1113903}{doi:\nolinkurl{10.3389/fcomp.2023.1113903}}


\bibitem[Buller and Burgoon(1994)]%
        {buller_deception_1994}
\bibfield{author}{\bibinfo{person}{David~B. Buller} {and} \bibinfo{person}{Judee~K. Burgoon}.} \bibinfo{year}{1994}\natexlab{}.
\newblock \showarticletitle{Deception: Strategic and Nonstrategic Communication}.
\newblock In \bibinfo{booktitle}{\emph{Strategic Interpersonal Communication}}. \bibinfo{publisher}{Routledge}.
\newblock
\newblock
\shownote{Num Pages: 33}.


\bibitem[Buller and Burgoon(1996)]%
        {buller_interpersonal_1996}
\bibfield{author}{\bibinfo{person}{David~B. Buller} {and} \bibinfo{person}{Judee~K. Burgoon}.} \bibinfo{year}{1996}\natexlab{}.
\newblock \showarticletitle{Interpersonal Deception Theory}.
\newblock \bibinfo{journal}{\emph{Communication Theory}} \bibinfo{volume}{6}, \bibinfo{number}{3} (\bibinfo{year}{1996}), \bibinfo{pages}{203--242}.
\newblock
\showISSN{1050-3293, 1468-2885}
\href{https://doi.org/10.1111/j.1468-2885.1996.tb00127.x}{doi:\nolinkurl{10.1111/j.1468-2885.1996.tb00127.x}}


\bibitem[Burleson et~al\mbox{.}(1987)]%
        {burleson_cognitive_1987}
\bibfield{author}{\bibinfo{person}{Brant~R. Burleson}, \bibinfo{person}{J.~C. {McCroskey}}, {and} \bibinfo{person}{J.~A. Daly}.} \bibinfo{year}{1987}\natexlab{}.
\newblock \showarticletitle{Cognitive complexity}.
\newblock \bibinfo{journal}{\emph{Personality and interpersonal communication}} (\bibinfo{year}{1987}), \bibinfo{pages}{305--349}.
\newblock


\bibitem[Cai et~al\mbox{.}(2024)]%
        {cai_antagonistic_2024}
\bibfield{author}{\bibinfo{person}{Alice Cai}, \bibinfo{person}{Ian Arawjo}, {and} \bibinfo{person}{Elena~L. Glassman}.} \bibinfo{year}{2024}\natexlab{}.
\newblock \bibinfo{title}{Antagonistic {AI}}.
\newblock
\showeprint[arxiv]{2402.07350 [cs]}
\href{https://doi.org/10.48550/arXiv.2402.07350}{doi:\nolinkurl{10.48550/arXiv.2402.07350}}


\bibitem[Carro(2024)]%
        {carro_flattering_2024}
\bibfield{author}{\bibinfo{person}{María~Victoria Carro}.} \bibinfo{year}{2024}\natexlab{}.
\newblock \bibinfo{title}{Flattering to Deceive: The Impact of Sycophantic Behavior on User Trust in Large Language Model}.
\newblock
\href{https://doi.org/10.48550/ARXIV.2412.02802}{doi:\nolinkurl{10.48550/ARXIV.2412.02802}}


\bibitem[Cavazza(2016)]%
        {cavazza_when_2016}
\bibfield{author}{\bibinfo{person}{Nicoletta Cavazza}.} \bibinfo{year}{2016}\natexlab{}.
\newblock \showarticletitle{When political candidates “go positive”: the effects of flattering the rival in political communication}.
\newblock \bibinfo{journal}{\emph{Social Influence}} \bibinfo{volume}{11}, \bibinfo{number}{3} (\bibinfo{year}{2016}), \bibinfo{pages}{166--176}.
\newblock
\showISSN{1553-4510, 1553-4529}
\href{https://doi.org/10.1080/15534510.2016.1206962}{doi:\nolinkurl{10.1080/15534510.2016.1206962}}


\bibitem[Cavazza and Guidetti(2018)]%
        {cavazza_captatio_2018}
\bibfield{author}{\bibinfo{person}{Nicoletta Cavazza} {and} \bibinfo{person}{Margherita Guidetti}.} \bibinfo{year}{2018}\natexlab{}.
\newblock \showarticletitle{\textit{Captatio Benevolentiae} : Potential Risks and Benefits of Flattering the Audience in a Public Political Speech}.
\newblock \bibinfo{journal}{\emph{Journal of Language and Social Psychology}} \bibinfo{volume}{37}, \bibinfo{number}{6} (\bibinfo{year}{2018}), \bibinfo{pages}{706--720}.
\newblock
\showISSN{0261-927X, 1552-6526}
\href{https://doi.org/10.1177/0261927X18800132}{doi:\nolinkurl{10.1177/0261927X18800132}}


\bibitem[{CBC News}(2025)]%
        {cbc2025aipsychosis}
\bibfield{author}{\bibinfo{person}{{CBC News}}.} \bibinfo{year}{2025}\natexlab{}.
\newblock \bibinfo{booktitle}{\emph{AI psychosis: Long talks with chatbots left these men with ‘AI psychosis’}}.
\newblock
\urldef\tempurl%
\url{https://www.cbc.ca/news/canada/ai-psychosis-canada-1.7631925}
\showURL{%
\tempurl}
\newblock
\shownote{CBC}.


\bibitem[Chan and Sengupta(2010)]%
        {chan_insincere_2010}
\bibfield{author}{\bibinfo{person}{Elaine Chan} {and} \bibinfo{person}{Jaideep Sengupta}.} \bibinfo{year}{2010}\natexlab{}.
\newblock \showarticletitle{Insincere Flattery Actually Works: A Dual Attitudes Perspective}.
\newblock \bibinfo{journal}{\emph{Journal of Marketing Research}} \bibinfo{volume}{47}, \bibinfo{number}{1} (\bibinfo{year}{2010}), \bibinfo{pages}{122--133}.
\newblock
\showISSN{0022-2437, 1547-7193}
\href{https://doi.org/10.1509/jmkr.47.1.122}{doi:\nolinkurl{10.1509/jmkr.47.1.122}}


\bibitem[Chan and Sengupta(2013)]%
        {chan_observing_2013}
\bibfield{author}{\bibinfo{person}{Elaine Chan} {and} \bibinfo{person}{Jaideep Sengupta}.} \bibinfo{year}{2013}\natexlab{}.
\newblock \showarticletitle{Observing Flattery: A Social Comparison Perspective}.
\newblock \bibinfo{journal}{\emph{Journal of Consumer Research}} \bibinfo{volume}{40}, \bibinfo{number}{4} (\bibinfo{year}{2013}), \bibinfo{pages}{740--758}.
\newblock
\showISSN{0093-5301, 1537-5277}
\href{https://doi.org/10.1086/672357}{doi:\nolinkurl{10.1086/672357}}


\bibitem[Chen et~al\mbox{.}(2024)]%
        {chen_is_2024}
\bibfield{author}{\bibinfo{person}{Cheng Chen}, \bibinfo{person}{Sangwook Lee}, \bibinfo{person}{Eunchae Jang}, {and} \bibinfo{person}{S.~Shyam Sundar}.} \bibinfo{year}{2024}\natexlab{}.
\newblock \showarticletitle{Is Your Prompt Detailed Enough? Exploring the Effects of Prompt Coaching on Users' Perceptions, Engagement, and Trust in Text-to-Image Generative {AI} Tools}. In \bibinfo{booktitle}{\emph{Proceedings of the Second International Symposium on Trustworthy Autonomous Systems}} (New York, {NY}, {USA}, 2024-09-16) \emph{(\bibinfo{series}{{TAS} '24})}. \bibinfo{publisher}{Association for Computing Machinery}, \bibinfo{pages}{1--12}.
\newblock
\showISBNx{979-8-4007-0989-0}
\href{https://doi.org/10.1145/3686038.3686060}{doi:\nolinkurl{10.1145/3686038.3686060}}


\bibitem[Cheng et~al\mbox{.}(2025)]%
        {cheng_social_2025}
\bibfield{author}{\bibinfo{person}{Myra Cheng}, \bibinfo{person}{Sunny Yu}, \bibinfo{person}{Cinoo Lee}, \bibinfo{person}{Pranav Khadpe}, \bibinfo{person}{Lujain Ibrahim}, {and} \bibinfo{person}{Dan Jurafsky}.} \bibinfo{year}{2025}\natexlab{}.
\newblock \bibinfo{title}{Social Sycophancy: A Broader Understanding of {LLM} Sycophancy}.
\newblock
\href{https://doi.org/10.48550/ARXIV.2505.13995}{doi:\nolinkurl{10.48550/ARXIV.2505.13995}}


\bibitem[Chiriatti et~al\mbox{.}(2025)]%
        {chiriatti_system_2025}
\bibfield{author}{\bibinfo{person}{Massimo Chiriatti}, \bibinfo{person}{Marianna Bergamaschi~Ganapini}, \bibinfo{person}{Enrico Panai}, \bibinfo{person}{Brenda~K. Wiederhold}, {and} \bibinfo{person}{Giuseppe Riva}.} \bibinfo{year}{2025}\natexlab{}.
\newblock \showarticletitle{System 0: Transforming Artificial Intelligence into a Cognitive Extension}.
\newblock \bibinfo{journal}{\emph{Cyberpsychology, Behavior, and Social Networking}} \bibinfo{volume}{28}, \bibinfo{number}{7} (\bibinfo{year}{2025}), \bibinfo{pages}{534--542}.
\newblock
\showISSN{2152-2715, 2152-2723}
\href{https://doi.org/10.1089/cyber.2025.0201}{doi:\nolinkurl{10.1089/cyber.2025.0201}}


\bibitem[Chisholm and Feehan(1977)]%
        {chisholm_intent_1977}
\bibfield{author}{\bibinfo{person}{Roderick~M. Chisholm} {and} \bibinfo{person}{Thomas~D. Feehan}.} \bibinfo{year}{1977}\natexlab{}.
\newblock \showarticletitle{The Intent to Deceive}.
\newblock \bibinfo{journal}{\emph{The Journal of Philosophy}} \bibinfo{volume}{74}, \bibinfo{number}{3} (\bibinfo{year}{1977}), \bibinfo{pages}{143}.
\newblock
\showISSN{0022362X}
\href{https://doi.org/10.2307/2025605}{doi:\nolinkurl{10.2307/2025605}}


\bibitem[Chu et~al\mbox{.}(2025)]%
        {chu_illusions_2025}
\bibfield{author}{\bibinfo{person}{Minh~Duc Chu}, \bibinfo{person}{Patrick Gerard}, \bibinfo{person}{Kshitij Pawar}, \bibinfo{person}{Charles Bickham}, {and} \bibinfo{person}{Kristina Lerman}.} \bibinfo{year}{2025}\natexlab{}.
\newblock \bibinfo{title}{Illusions of Intimacy: Emotional Attachment and Emerging Psychological Risks in Human-{AI} Relationships}.
\newblock
\showeprint[arxiv]{2505.11649 [cs]}
\href{https://doi.org/10.48550/arXiv.2505.11649}{doi:\nolinkurl{10.48550/arXiv.2505.11649}}


\bibitem[Clarke et~al\mbox{.}(2022)]%
        {clarke_effects_2022}
\bibfield{author}{\bibinfo{person}{Nicholas Clarke}, \bibinfo{person}{Najla Alshenaifi}, {and} \bibinfo{person}{Thomas Garavan}.} \bibinfo{year}{2022}\natexlab{}.
\newblock \showarticletitle{The effects of subordinates’ use of upward influence tactics on their supervisors’ job performance evaluations in Saudi Arabia: the significance of loyalty}.
\newblock \bibinfo{journal}{\emph{The International Journal of Human Resource Management}} \bibinfo{volume}{33}, \bibinfo{number}{2} (\bibinfo{year}{2022}), \bibinfo{pages}{239--268}.
\newblock
\showISSN{0958-5192}
\href{https://doi.org/10.1080/09585192.2019.1686650}{doi:\nolinkurl{10.1080/09585192.2019.1686650}}


\bibitem[Colman and Olver(1978)]%
        {colman_reactions_1978}
\bibfield{author}{\bibinfo{person}{Andrew~M. Colman} {and} \bibinfo{person}{Kevin~R. Olver}.} \bibinfo{year}{1978}\natexlab{}.
\newblock \showarticletitle{Reactions to flattery as a function of self‐esteem: Self‐enhancement and cognitive consistency theories}.
\newblock \bibinfo{journal}{\emph{British Journal of Social and Clinical Psychology}} \bibinfo{volume}{17}, \bibinfo{number}{1} (\bibinfo{year}{1978}), \bibinfo{pages}{25--29}.
\newblock
\showISSN{0007-1293}
\href{https://doi.org/10.1111/j.2044-8260.1978.tb00892.x}{doi:\nolinkurl{10.1111/j.2044-8260.1978.tb00892.x}}


\bibitem[Daft et~al\mbox{.}(1987)]%
        {daft_message_1987}
\bibfield{author}{\bibinfo{person}{Richard~L. Daft}, \bibinfo{person}{Robert~H. Lengel}, {and} \bibinfo{person}{Linda~Klebe Trevino}.} \bibinfo{year}{1987}\natexlab{}.
\newblock \showarticletitle{Message Equivocality, Media Selection, and Manager Performance: Implications for Information Systems}.
\newblock \bibinfo{journal}{\emph{{MIS} Quarterly}} \bibinfo{volume}{11}, \bibinfo{number}{3} (\bibinfo{year}{1987}), \bibinfo{pages}{355}.
\newblock
\showISSN{02767783}
\href{https://doi.org/10.2307/248682}{doi:\nolinkurl{10.2307/248682}}


\bibitem[Danziger(2020)]%
        {danziger_pragmatics_2020}
\bibfield{author}{\bibinfo{person}{Roni Danziger}.} \bibinfo{year}{2020}\natexlab{}.
\newblock \showarticletitle{The pragmatics of flattery: The strategic use of solidarity-oriented actions}.
\newblock \bibinfo{journal}{\emph{Journal of Pragmatics}}  \bibinfo{volume}{170} (\bibinfo{year}{2020}), \bibinfo{pages}{413--425}.
\newblock
\showISSN{03782166}
\href{https://doi.org/10.1016/j.pragma.2020.09.027}{doi:\nolinkurl{10.1016/j.pragma.2020.09.027}}


\bibitem[Danziger(2021)]%
        {danziger_democratic_2021}
\bibfield{author}{\bibinfo{person}{Roni Danziger}.} \bibinfo{year}{2021}\natexlab{}.
\newblock \showarticletitle{The democratic king: The role of ritualized flattery in political discourse}.
\newblock \bibinfo{journal}{\emph{Discourse \& Society}} \bibinfo{volume}{32}, \bibinfo{number}{6} (\bibinfo{year}{2021}), \bibinfo{pages}{645--665}.
\newblock
\showISSN{0957-9265, 1460-3624}
\href{https://doi.org/10.1177/09579265211023224}{doi:\nolinkurl{10.1177/09579265211023224}}


\bibitem[Diken(2025)]%
        {diken2025flattery}
\bibfield{author}{\bibinfo{person}{B{\"u}lent Diken}.} \bibinfo{year}{2025}\natexlab{}.
\newblock \showarticletitle{Flattery, Truth-telling, and Social Theory}.
\newblock \bibinfo{journal}{\emph{Theory, Culture \& Society}} \bibinfo{volume}{42}, \bibinfo{number}{4} (\bibinfo{year}{2025}), \bibinfo{pages}{02632764251323864}.
\newblock
\href{https://doi.org/10.1177/02632764251323864}{doi:\nolinkurl{10.1177/02632764251323864}}


\bibitem[Ding et~al\mbox{.}(2023)]%
        {ding_students_2023}
\bibfield{author}{\bibinfo{person}{Lu Ding}, \bibinfo{person}{Tong Li}, \bibinfo{person}{Shiyan Jiang}, {and} \bibinfo{person}{Albert Gapud}.} \bibinfo{year}{2023}\natexlab{}.
\newblock \showarticletitle{Students’ perceptions of using {ChatGPT} in a physics class as a virtual tutor}.
\newblock \bibinfo{journal}{\emph{International Journal of Educational Technology in Higher Education}} \bibinfo{volume}{20}, \bibinfo{number}{1} (\bibinfo{year}{2023}), \bibinfo{pages}{63}.
\newblock
\showISSN{2365-9440}
\href{https://doi.org/10.1186/s41239-023-00434-1}{doi:\nolinkurl{10.1186/s41239-023-00434-1}}


\bibitem[Dohnány et~al\mbox{.}(2025)]%
        {dohnany_technological_2025}
\bibfield{author}{\bibinfo{person}{Sebastian Dohnány}, \bibinfo{person}{Zeb Kurth-Nelson}, \bibinfo{person}{Eleanor Spens}, \bibinfo{person}{Lennart Luettgau}, \bibinfo{person}{Alastair Reid}, \bibinfo{person}{Iason Gabriel}, \bibinfo{person}{Christopher Summerfield}, \bibinfo{person}{Murray Shanahan}, {and} \bibinfo{person}{Matthew~M. Nour}.} \bibinfo{year}{2025}\natexlab{}.
\newblock \bibinfo{title}{Technological folie à deux: Feedback Loops Between {AI} Chatbots and Mental Illness}.
\newblock
\showeprint[arxiv]{2507.19218 [cs]}
\href{https://doi.org/10.48550/arXiv.2507.19218}{doi:\nolinkurl{10.48550/arXiv.2507.19218}}


\bibitem[Eagly and Chaiken(1998)]%
        {eagly_attitude_1998}
\bibfield{author}{\bibinfo{person}{Alice~H. Eagly} {and} \bibinfo{person}{Shelly Chaiken}.} \bibinfo{year}{1998}\natexlab{}.
\newblock \showarticletitle{Attitude structure and function}.
\newblock In \bibinfo{booktitle}{\emph{The handbook of social psychology, Vols. 1-2, 4th ed}}. \bibinfo{publisher}{{McGraw}-Hill}, \bibinfo{pages}{269--322}.
\newblock
\showISBNx{978-0-19-521376-8}


\bibitem[Eylon and Heyd(2008)]%
        {eylon_flattery_2008}
\bibfield{author}{\bibinfo{person}{Yuval Eylon} {and} \bibinfo{person}{David Heyd}.} \bibinfo{year}{2008}\natexlab{}.
\newblock \showarticletitle{Flattery}.
\newblock \bibinfo{journal}{\emph{Philosophy and Phenomenological Research}} \bibinfo{volume}{77}, \bibinfo{number}{3} (\bibinfo{year}{2008}), \bibinfo{pages}{685--704}.
\newblock
\showISSN{0031-8205, 1933-1592}
\href{https://doi.org/10.1111/j.1933-1592.2008.00215.x}{doi:\nolinkurl{10.1111/j.1933-1592.2008.00215.x}}


\bibitem[Fan et~al\mbox{.}(2025)]%
        {fan_fairmt-bench_2025}
\bibfield{author}{\bibinfo{person}{Zhiting Fan}, \bibinfo{person}{Ruizhe Chen}, \bibinfo{person}{Tianxiang Hu}, {and} \bibinfo{person}{Zuozhu Liu}.} \bibinfo{year}{2025}\natexlab{}.
\newblock \bibinfo{title}{{FairMT}-Bench: Benchmarking Fairness for Multi-turn Dialogue in Conversational {LLMs}}.
\newblock
\showeprint[arxiv]{2410.19317 [cs]}
\href{https://doi.org/10.48550/arXiv.2410.19317}{doi:\nolinkurl{10.48550/arXiv.2410.19317}}


\bibitem[Fanous et~al\mbox{.}(2025)]%
        {fanous_syceval_2025}
\bibfield{author}{\bibinfo{person}{Aaron Fanous}, \bibinfo{person}{Jacob Goldberg}, \bibinfo{person}{Ank~A. Agarwal}, \bibinfo{person}{Joanna Lin}, \bibinfo{person}{Anson Zhou}, \bibinfo{person}{Roxana Daneshjou}, {and} \bibinfo{person}{Sanmi Koyejo}.} \bibinfo{year}{2025}\natexlab{}.
\newblock \bibinfo{title}{{SycEval}: Evaluating {LLM} Sycophancy}.
\newblock
\href{https://doi.org/10.48550/ARXIV.2502.08177}{doi:\nolinkurl{10.48550/ARXIV.2502.08177}}


\bibitem[Ferino et~al\mbox{.}(2025)]%
        {ferino_novice_2025}
\bibfield{author}{\bibinfo{person}{Samuel Ferino}, \bibinfo{person}{Rashina Hoda}, \bibinfo{person}{John Grundy}, {and} \bibinfo{person}{Christoph Treude}.} \bibinfo{year}{2025}\natexlab{}.
\newblock \bibinfo{title}{Novice Developers' Perspectives on Adopting {LLMs} for Software Development: A Systematic Literature Review}.
\newblock
\href{https://doi.org/10.48550/ARXIV.2503.07556}{doi:\nolinkurl{10.48550/ARXIV.2503.07556}}
\newblock
\shownote{Version Number: 2}.


\bibitem[Fogg and Nass(1997)]%
        {fogg_silicon_1997}
\bibfield{author}{\bibinfo{person}{B.J. Fogg} {and} \bibinfo{person}{Clifford Nass}.} \bibinfo{year}{1997}\natexlab{}.
\newblock \showarticletitle{Silicon sycophants: the effects of computers that flatter}.
\newblock \bibinfo{journal}{\emph{International Journal of Human-Computer Studies}} \bibinfo{volume}{46}, \bibinfo{number}{5} (\bibinfo{year}{1997}), \bibinfo{pages}{551--561}.
\newblock
\showISSN{10715819}
\href{https://doi.org/10.1006/ijhc.1996.0104}{doi:\nolinkurl{10.1006/ijhc.1996.0104}}


\bibitem[Goffman(1955)]%
        {goffman_face-work_1955}
\bibfield{author}{\bibinfo{person}{Erving Goffman}.} \bibinfo{year}{1955}\natexlab{}.
\newblock \showarticletitle{On Face-Work: An Analysis of Ritual Elements in Social Interaction}.
\newblock \bibinfo{journal}{\emph{Psychiatry}} \bibinfo{volume}{18}, \bibinfo{number}{3} (\bibinfo{year}{1955}), \bibinfo{pages}{213--231}.
\newblock
\showISSN{0033-2747, 1943-281X}
\href{https://doi.org/10.1080/00332747.1955.11023008}{doi:\nolinkurl{10.1080/00332747.1955.11023008}}


\bibitem[Hancock et~al\mbox{.}(2011)]%
        {hancock_meta-analysis_2011}
\bibfield{author}{\bibinfo{person}{Peter~A. Hancock}, \bibinfo{person}{Deborah~R. Billings}, \bibinfo{person}{Kristin~E. Schaefer}, \bibinfo{person}{Jessie Y.~C. Chen}, \bibinfo{person}{Ewart~J. De~Visser}, {and} \bibinfo{person}{Raja Parasuraman}.} \bibinfo{year}{2011}\natexlab{}.
\newblock \showarticletitle{A Meta-Analysis of Factors Affecting Trust in Human-Robot Interaction}.
\newblock \bibinfo{journal}{\emph{Human Factors: The Journal of the Human Factors and Ergonomics Society}} \bibinfo{volume}{53}, \bibinfo{number}{5} (\bibinfo{year}{2011}), \bibinfo{pages}{517--527}.
\newblock
\showISSN{0018-7208, 1547-8181}
\href{https://doi.org/10.1177/0018720811417254}{doi:\nolinkurl{10.1177/0018720811417254}}


\bibitem[High and Dillard(2012)]%
        {high_review_2012}
\bibfield{author}{\bibinfo{person}{Andrew~C. High} {and} \bibinfo{person}{James~Price Dillard}.} \bibinfo{year}{2012}\natexlab{}.
\newblock \showarticletitle{A Review and Meta-Analysis of Person-Centered Messages and Social Support Outcomes}.
\newblock \bibinfo{journal}{\emph{Communication Studies}} \bibinfo{volume}{63}, \bibinfo{number}{1} (\bibinfo{year}{2012}), \bibinfo{pages}{99--118}.
\newblock
\showISSN{1051-0974, 1745-1035}
\href{https://doi.org/10.1080/10510974.2011.598208}{doi:\nolinkurl{10.1080/10510974.2011.598208}}


\bibitem[Huang et~al\mbox{.}(2025)]%
        {huang_survey_2025}
\bibfield{author}{\bibinfo{person}{Lei Huang}, \bibinfo{person}{Weijiang Yu}, \bibinfo{person}{Weitao Ma}, \bibinfo{person}{Weihong Zhong}, \bibinfo{person}{Zhangyin Feng}, \bibinfo{person}{Haotian Wang}, \bibinfo{person}{Qianglong Chen}, \bibinfo{person}{Weihua Peng}, \bibinfo{person}{Xiaocheng Feng}, \bibinfo{person}{Bing Qin}, {and} \bibinfo{person}{Ting Liu}.} \bibinfo{year}{2025}\natexlab{}.
\newblock \showarticletitle{A Survey on Hallucination in Large Language Models: Principles, Taxonomy, Challenges, and Open Questions}.
\newblock \bibinfo{journal}{\emph{{ACM} Transactions on Information Systems}} \bibinfo{volume}{43}, \bibinfo{number}{2} (\bibinfo{year}{2025}), \bibinfo{pages}{1--55}.
\newblock
\showISSN{1046-8188, 1558-2868}
\href{https://doi.org/10.1145/3703155}{doi:\nolinkurl{10.1145/3703155}}


\bibitem[Håkansson~Eklund and Summer~Meranius(2021)]%
        {hakansson_eklund_toward_2021}
\bibfield{author}{\bibinfo{person}{Jakob Håkansson~Eklund} {and} \bibinfo{person}{Martina Summer~Meranius}.} \bibinfo{year}{2021}\natexlab{}.
\newblock \showarticletitle{Toward a consensus on the nature of empathy: A review of reviews}.
\newblock \bibinfo{journal}{\emph{Patient Education and Counseling}} \bibinfo{volume}{104}, \bibinfo{number}{2} (\bibinfo{year}{2021}), \bibinfo{pages}{300--307}.
\newblock
\showISSN{0738-3991}
\href{https://doi.org/10.1016/j.pec.2020.08.022}{doi:\nolinkurl{10.1016/j.pec.2020.08.022}}


\bibitem[Jami et~al\mbox{.}(2024)]%
        {jami_interaction_2024}
\bibfield{author}{\bibinfo{person}{Parvaneh~Yaghoubi Jami}, \bibinfo{person}{David~Ian Walker}, {and} \bibinfo{person}{Behzad Mansouri}.} \bibinfo{year}{2024}\natexlab{}.
\newblock \showarticletitle{Interaction of empathy and culture: a review}.
\newblock \bibinfo{journal}{\emph{Current Psychology}} \bibinfo{volume}{43}, \bibinfo{number}{4} (\bibinfo{year}{2024}), \bibinfo{pages}{2965--2980}.
\newblock
\showISSN{1046-1310, 1936-4733}
\href{https://doi.org/10.1007/s12144-023-04422-6}{doi:\nolinkurl{10.1007/s12144-023-04422-6}}


\bibitem[Johnson et~al\mbox{.}(2004)]%
        {johnson_experience_2004}
\bibfield{author}{\bibinfo{person}{Daniel Johnson}, \bibinfo{person}{John Gardner}, {and} \bibinfo{person}{Janet Wiles}.} \bibinfo{year}{2004}\natexlab{}.
\newblock \showarticletitle{Experience as a moderator of the media equation: the impact of flattery and praise}.
\newblock \bibinfo{journal}{\emph{International Journal of Human-Computer Studies}} \bibinfo{volume}{61}, \bibinfo{number}{3} (\bibinfo{year}{2004}), \bibinfo{pages}{237--258}.
\newblock
\showISSN{1071-5819}
\href{https://doi.org/10.1016/j.ijhcs.2003.12.008}{doi:\nolinkurl{10.1016/j.ijhcs.2003.12.008}}


\bibitem[Kang and Lou(2022)]%
        {kang_ai_2022}
\bibfield{author}{\bibinfo{person}{Hyunjin Kang} {and} \bibinfo{person}{Chen Lou}.} \bibinfo{year}{2022}\natexlab{}.
\newblock \showarticletitle{{AI} agency vs. human agency: understanding human–{AI} interactions on {TikTok} and their implications for user engagement}.
\newblock \bibinfo{journal}{\emph{Journal of Computer-Mediated Communication}} \bibinfo{volume}{27}, \bibinfo{number}{5} (\bibinfo{year}{2022}), \bibinfo{pages}{zmac014}.
\newblock
\showISSN{1083-6101}
\href{https://doi.org/10.1093/jcmc/zmac014}{doi:\nolinkurl{10.1093/jcmc/zmac014}}


\bibitem[Khalifa(2022)]%
        {khalifa_conceptual_2022}
\bibfield{author}{\bibinfo{person}{Hussein Khalifa~Hassan Khalifa}.} \bibinfo{year}{2022}\natexlab{}.
\newblock \showarticletitle{A Conceptual Review on Heuristic Systematic Model in Mass Communication Studies}.
\newblock \bibinfo{journal}{\emph{International Journal of Media and Mass Communication}} \bibinfo{volume}{04}, \bibinfo{number}{2} (\bibinfo{year}{2022}), \bibinfo{pages}{164--175--}.
\newblock
\showISSN{2535-9797}
\href{https://doi.org/10.46988/IJMMC.04.02.2022.007}{doi:\nolinkurl{10.46988/IJMMC.04.02.2022.007}}


\bibitem[Klawitter and Hargittai(2018)]%
        {klawitter_shortcuts_2018}
\bibfield{author}{\bibinfo{person}{Erin Klawitter} {and} \bibinfo{person}{Eszter Hargittai}.} \bibinfo{year}{2018}\natexlab{}.
\newblock \showarticletitle{Shortcuts to Well Being? Evaluating the Credibility of Online Health Information through Multiple Complementary Heuristics}.
\newblock \bibinfo{journal}{\emph{Journal of Broadcasting \& Electronic Media}} \bibinfo{volume}{62}, \bibinfo{number}{2} (\bibinfo{year}{2018}), \bibinfo{pages}{251--268}.
\newblock
\showISSN{0883-8151, 1550-6878}
\href{https://doi.org/10.1080/08838151.2018.1451863}{doi:\nolinkurl{10.1080/08838151.2018.1451863}}


\bibitem[Kong et~al\mbox{.}(2025)]%
        {kong_sharp_2025}
\bibfield{author}{\bibinfo{person}{Chuyi Kong}, \bibinfo{person}{Ziyang Luo}, \bibinfo{person}{Hongzhan Lin}, \bibinfo{person}{Zhiyuan Fan}, \bibinfo{person}{Yaxin Fan}, \bibinfo{person}{Yuxi Sun}, {and} \bibinfo{person}{Jing Ma}.} \bibinfo{year}{2025}\natexlab{}.
\newblock \bibinfo{title}{{SHARP}: {Unlocking} {Interactive} {Hallucination} via {Stance} {Transfer} in {Role}-{Playing} {LLMs}}.
\newblock
\href{https://doi.org/10.48550/arXiv.2411.07965}{doi:\nolinkurl{10.48550/arXiv.2411.07965}}
\newblock
\shownote{arXiv:2411.07965 [cs]}.


\bibitem[Kran et~al\mbox{.}(2025)]%
        {kran_darkbench_2025}
\bibfield{author}{\bibinfo{person}{Esben Kran}, \bibinfo{person}{Hieu Minh~"Jord" Nguyen}, \bibinfo{person}{Akash Kundu}, \bibinfo{person}{Sami Jawhar}, \bibinfo{person}{Jinsuk Park}, {and} \bibinfo{person}{Mateusz~Maria Jurewicz}.} \bibinfo{year}{2025}\natexlab{}.
\newblock \bibinfo{title}{{DarkBench}: Benchmarking Dark Patterns in Large Language Models}.
\newblock
\showeprint[arxiv]{2503.10728 [cs]}
\href{https://doi.org/10.48550/arXiv.2503.10728}{doi:\nolinkurl{10.48550/arXiv.2503.10728}}


\bibitem[Laestadius et~al\mbox{.}(2024)]%
        {laestadius_too_2024}
\bibfield{author}{\bibinfo{person}{Linnea Laestadius}, \bibinfo{person}{Andrea Bishop}, \bibinfo{person}{Michael Gonzalez}, \bibinfo{person}{Diana Illenčík}, {and} \bibinfo{person}{Celeste Campos-Castillo}.} \bibinfo{year}{2024}\natexlab{}.
\newblock \showarticletitle{Too human and not human enough: A grounded theory analysis of mental health harms from emotional dependence on the social chatbot Replika}.
\newblock \bibinfo{journal}{\emph{New Media \& Society}} \bibinfo{volume}{26}, \bibinfo{number}{10} (\bibinfo{year}{2024}), \bibinfo{pages}{5923--5941}.
\newblock
\showISSN{1461-4448, 1461-7315}
\href{https://doi.org/10.1177/14614448221142007}{doi:\nolinkurl{10.1177/14614448221142007}}


\bibitem[Lee(2008)]%
        {lee_flattery_2008}
\bibfield{author}{\bibinfo{person}{Eun-Ju Lee}.} \bibinfo{year}{2008}\natexlab{}.
\newblock \showarticletitle{Flattery may get computers somewhere, sometimes: The moderating role of output modality, computer gender, and user gender}.
\newblock \bibinfo{journal}{\emph{International Journal of Human-Computer Studies}} \bibinfo{volume}{66}, \bibinfo{number}{11} (\bibinfo{year}{2008}), \bibinfo{pages}{789--800}.
\newblock
\showISSN{1071-5819}
\href{https://doi.org/10.1016/j.ijhcs.2008.07.009}{doi:\nolinkurl{10.1016/j.ijhcs.2008.07.009}}


\bibitem[Lee(2009)]%
        {lee_i_2009}
\bibfield{author}{\bibinfo{person}{Eun-Ju Lee}.} \bibinfo{year}{2009}\natexlab{}.
\newblock \showarticletitle{I like you, but I won’t listen to you: Effects of rationality on affective and behavioral responses to computers that flatter}.
\newblock \bibinfo{journal}{\emph{International Journal of Human-Computer Studies}} \bibinfo{volume}{67}, \bibinfo{number}{8} (\bibinfo{year}{2009}), \bibinfo{pages}{628--638}.
\newblock
\showISSN{10715819}
\href{https://doi.org/10.1016/j.ijhcs.2009.03.003}{doi:\nolinkurl{10.1016/j.ijhcs.2009.03.003}}


\bibitem[Lee(2010a)]%
        {lee_more_2010}
\bibfield{author}{\bibinfo{person}{Eun-Ju Lee}.} \bibinfo{year}{2010}\natexlab{a}.
\newblock \showarticletitle{The more humanlike, the better? How speech type and users’ cognitive style affect social responses to computers}.
\newblock \bibinfo{journal}{\emph{Computers in Human Behavior}} \bibinfo{volume}{26}, \bibinfo{number}{4} (\bibinfo{year}{2010}), \bibinfo{pages}{665--672}.
\newblock
\showISSN{0747-5632}
\href{https://doi.org/10.1016/j.chb.2010.01.003}{doi:\nolinkurl{10.1016/j.chb.2010.01.003}}


\bibitem[Lee(2010b)]%
        {lee_what_2010}
\bibfield{author}{\bibinfo{person}{Eun-Ju Lee}.} \bibinfo{year}{2010}\natexlab{b}.
\newblock \showarticletitle{What Triggers Social Responses to Flattering Computers? Experimental Tests of Anthropomorphism and Mindlessness Explanations}.
\newblock \bibinfo{journal}{\emph{Communication Research}} \bibinfo{volume}{37}, \bibinfo{number}{2} (\bibinfo{year}{2010}), \bibinfo{pages}{191--214}.
\newblock
\showISSN{0093-6502, 1552-3810}
\href{https://doi.org/10.1177/0093650209356389}{doi:\nolinkurl{10.1177/0093650209356389}}


\bibitem[Lee(2024)]%
        {lee_minding_2024}
\bibfield{author}{\bibinfo{person}{Eun-Ju Lee}.} \bibinfo{year}{2024}\natexlab{}.
\newblock \showarticletitle{Minding the source: toward an integrative theory of human–machine communication}.
\newblock \bibinfo{journal}{\emph{Human Communication Research}} \bibinfo{volume}{50}, \bibinfo{number}{2} (\bibinfo{year}{2024}), \bibinfo{pages}{184--193}.
\newblock
\showISSN{1468-2958}
\href{https://doi.org/10.1093/hcr/hqad034}{doi:\nolinkurl{10.1093/hcr/hqad034}}


\bibitem[Li and Zhang(2024)]%
        {li_finding_2024}
\bibfield{author}{\bibinfo{person}{Han Li} {and} \bibinfo{person}{Renwen Zhang}.} \bibinfo{year}{2024}\natexlab{}.
\newblock \showarticletitle{Finding love in algorithms: deciphering the emotional contexts of close encounters with {AI} chatbots}.
\newblock \bibinfo{journal}{\emph{Journal of Computer-Mediated Communication}} \bibinfo{volume}{29}, \bibinfo{number}{5} (\bibinfo{year}{2024}), \bibinfo{pages}{zmae015}.
\newblock
\showISSN{1083-6101}
\href{https://doi.org/10.1093/jcmc/zmae015}{doi:\nolinkurl{10.1093/jcmc/zmae015}}


\bibitem[Li et~al\mbox{.}(2025)]%
        {li_beyond_2025}
\bibfield{author}{\bibinfo{person}{Yubo Li}, \bibinfo{person}{Xiaobin Shen}, \bibinfo{person}{Xinyu Yao}, \bibinfo{person}{Xueying Ding}, \bibinfo{person}{Yidi Miao}, \bibinfo{person}{Ramayya Krishnan}, {and} \bibinfo{person}{Rema Padman}.} \bibinfo{year}{2025}\natexlab{}.
\newblock \bibinfo{title}{Beyond Single-Turn: A Survey on Multi-Turn Interactions with Large Language Models}.
\newblock
\showeprint[arxiv]{2504.04717 [cs]}
\href{https://doi.org/10.48550/arXiv.2504.04717}{doi:\nolinkurl{10.48550/arXiv.2504.04717}}


\bibitem[Liao and Sundar(2022)]%
        {liao_designing_2022}
\bibfield{author}{\bibinfo{person}{Q.Vera Liao} {and} \bibinfo{person}{S.~Shyam Sundar}.} \bibinfo{year}{2022}\natexlab{}.
\newblock \showarticletitle{Designing for Responsible Trust in {AI} Systems: A Communication Perspective}. In \bibinfo{booktitle}{\emph{2022 {ACM} Conference on Fairness Accountability and Transparency}} (Seoul Republic of Korea, 2022-06-21). \bibinfo{publisher}{{ACM}}, \bibinfo{pages}{1257--1268}.
\newblock
\showISBNx{978-1-4503-9352-2}
\href{https://doi.org/10.1145/3531146.3533182}{doi:\nolinkurl{10.1145/3531146.3533182}}


\bibitem[Liden and Mitchell(1988)]%
        {liden1988ingratiatory}
\bibfield{author}{\bibinfo{person}{Robert~C Liden} {and} \bibinfo{person}{Terence~R Mitchell}.} \bibinfo{year}{1988}\natexlab{}.
\newblock \showarticletitle{Ingratiatory behaviors in organizational settings}.
\newblock \bibinfo{journal}{\emph{Academy of management review}} \bibinfo{volume}{13}, \bibinfo{number}{4} (\bibinfo{year}{1988}), \bibinfo{pages}{572--587}.
\newblock


\bibitem[{MacGeorge} et~al\mbox{.}(2011)]%
        {macgeorge_supportive_2011}
\bibfield{author}{\bibinfo{person}{Erina~L. {MacGeorge}}, \bibinfo{person}{Bo Feng}, {and} \bibinfo{person}{Brant~R. Burleson}.} \bibinfo{year}{2011}\natexlab{}.
\newblock \showarticletitle{Supportive communication}.
\newblock \bibinfo{journal}{\emph{Handbook of interpersonal communication}}  \bibinfo{volume}{4} (\bibinfo{year}{2011}), \bibinfo{pages}{317--354}.
\newblock
\urldef\tempurl%
\url{https://books.google.com/books?hl=zh-CN&lr=&id=ZNF1AwAAQBAJ&oi=fnd&pg=PA317&dq=Supportive+Communication++Erina+L.+MacGeorge++Bo+Feng++Brant+R.+Burleson&ots=rpge7_utrF&sig=14eJB3F88LGE32VQnhGc5-QimtA}
\showURL{%
\tempurl}


\bibitem[Malmqvist(2024)]%
        {malmqvist_sycophancy_2024}
\bibfield{author}{\bibinfo{person}{Lars Malmqvist}.} \bibinfo{year}{2024}\natexlab{}.
\newblock \bibinfo{title}{Sycophancy in Large Language Models: Causes and Mitigations}.
\newblock
\showeprint[arxiv]{2411.15287 [cs]}
\href{https://doi.org/10.48550/arXiv.2411.15287}{doi:\nolinkurl{10.48550/arXiv.2411.15287}}


\bibitem[Marriott and Pitardi(2024)]%
        {marriott_one_2024}
\bibfield{author}{\bibinfo{person}{Hannah~R. Marriott} {and} \bibinfo{person}{Valentina Pitardi}.} \bibinfo{year}{2024}\natexlab{}.
\newblock \showarticletitle{One is the loneliest number… Two can be as bad as one. The influence of {AI} Friendship Apps on users' well-being and addiction}.
\newblock \bibinfo{journal}{\emph{Psychology \& Marketing}} \bibinfo{volume}{41}, \bibinfo{number}{1} (\bibinfo{year}{2024}), \bibinfo{pages}{86--101}.
\newblock
\showISSN{1520-6793}
\href{https://doi.org/10.1002/mar.21899}{doi:\nolinkurl{10.1002/mar.21899}}


\bibitem[Nass et~al\mbox{.}(1994)]%
        {nass1994computers}
\bibfield{author}{\bibinfo{person}{Clifford Nass}, \bibinfo{person}{Jonathan Steuer}, {and} \bibinfo{person}{Ellen~R Tauber}.} \bibinfo{year}{1994}\natexlab{}.
\newblock \showarticletitle{Computers are social actors}. In \bibinfo{booktitle}{\emph{Proceedings of the SIGCHI conference on Human factors in computing systems}}. \bibinfo{pages}{72--78}.
\newblock
\href{https://doi.org/10.1145/191666.191703}{doi:\nolinkurl{10.1145/191666.191703}}


\bibitem[Nickerson(1998)]%
        {nickerson1998confirmation}
\bibfield{author}{\bibinfo{person}{Raymond~S Nickerson}.} \bibinfo{year}{1998}\natexlab{}.
\newblock \showarticletitle{Confirmation bias: A ubiquitous phenomenon in many guises}.
\newblock \bibinfo{journal}{\emph{Review of general psychology}} \bibinfo{volume}{2}, \bibinfo{number}{2} (\bibinfo{year}{1998}), \bibinfo{pages}{175--220}.
\newblock
\href{https://doi.org/10.1037/1089-2680.2.2.175}{doi:\nolinkurl{10.1037/1089-2680.2.2.175}}


\bibitem[O’Sullivan and Carr(2018)]%
        {osullivan_masspersonal_2018}
\bibfield{author}{\bibinfo{person}{Patrick~B O’Sullivan} {and} \bibinfo{person}{Caleb~T Carr}.} \bibinfo{year}{2018}\natexlab{}.
\newblock \showarticletitle{Masspersonal communication: A model bridging the mass-interpersonal divide}.
\newblock \bibinfo{journal}{\emph{New Media \& Society}} \bibinfo{volume}{20}, \bibinfo{number}{3} (\bibinfo{year}{2018}), \bibinfo{pages}{1161--1180}.
\newblock
\showISSN{1461-4448, 1461-7315}
\href{https://doi.org/10.1177/1461444816686104}{doi:\nolinkurl{10.1177/1461444816686104}}


\bibitem[Panickssery(2023)]%
        {panickssery_reducing_2023}
\bibfield{author}{\bibinfo{person}{Nina Panickssery}.} \bibinfo{year}{2023}\natexlab{}.
\newblock \showarticletitle{Reducing sycophancy and improving honesty via activation steering}.
\newblock  (\bibinfo{year}{2023}).
\newblock
\urldef\tempurl%
\url{https://www.alignmentforum.org/posts/zt6hRsDE84HeBKh7E/reducing-sycophancy-and-improving-honesty-via-activation}
\showURL{%
\tempurl}


\bibitem[Park et~al\mbox{.}(2011)]%
        {park_set_2011}
\bibfield{author}{\bibinfo{person}{Sun~Hyun Park}, \bibinfo{person}{James~D. Westphal}, {and} \bibinfo{person}{Ithai Stern}.} \bibinfo{year}{2011}\natexlab{}.
\newblock \showarticletitle{Set up for a Fall: The Insidious Effects of Flattery and Opinion Conformity toward Corporate Leaders}.
\newblock \bibinfo{journal}{\emph{Administrative Science Quarterly}} \bibinfo{volume}{56}, \bibinfo{number}{2} (\bibinfo{year}{2011}), \bibinfo{pages}{257--302}.
\newblock
\showISSN{0001-8392, 1930-3815}
\href{https://doi.org/10.1177/0001839211429102}{doi:\nolinkurl{10.1177/0001839211429102}}


\bibitem[Paruchuri et~al\mbox{.}(2025)]%
        {paruchuri_whats_2025}
\bibfield{author}{\bibinfo{person}{Akshay Paruchuri}, \bibinfo{person}{Maryam Aziz}, \bibinfo{person}{Rohit Vartak}, \bibinfo{person}{Ayman Ali}, \bibinfo{person}{Best Uchehara}, \bibinfo{person}{Xin Liu}, \bibinfo{person}{Ishan Chatterjee}, {and} \bibinfo{person}{Monica Agrawal}.} \bibinfo{year}{2025}\natexlab{}.
\newblock \bibinfo{title}{"What's Up, Doc?": Analyzing How Users Seek Health Information in Large-Scale Conversational {AI} Datasets}.
\newblock
\href{https://doi.org/10.48550/ARXIV.2506.21532}{doi:\nolinkurl{10.48550/ARXIV.2506.21532}}


\bibitem[Patterson and Edinger(1987)]%
        {patterson1987functional}
\bibfield{author}{\bibinfo{person}{Miles~L. Patterson} {and} \bibinfo{person}{Joyce~A. Edinger}.} \bibinfo{year}{1987}\natexlab{}.
\newblock \bibinfo{booktitle}{\emph{A Functional Analysis of Space in Social Interaction}}.
\newblock \bibinfo{publisher}{Lawrence Erlbaum Associates}, \bibinfo{address}{Hillsdale, NJ}.
\newblock


\bibitem[Petty and Cacioppo(1986)]%
        {petty_message_1986}
\bibfield{author}{\bibinfo{person}{Richard~E. Petty} {and} \bibinfo{person}{John~T. Cacioppo}.} \bibinfo{year}{1986}\natexlab{}.
\newblock \showarticletitle{Message Elaboration versus Peripheral Cues}.
\newblock In \bibinfo{booktitle}{\emph{Communication and Persuasion: Central and Peripheral Routes to Attitude Change}}, \bibfield{editor}{\bibinfo{person}{Richard~E. Petty} {and} \bibinfo{person}{John~T. Cacioppo}} (Eds.). \bibinfo{publisher}{Springer}, \bibinfo{pages}{141--172}.
\newblock
\showISBNx{978-1-4612-4964-1}
\href{https://doi.org/10.1007/978-1-4612-4964-1_6}{doi:\nolinkurl{10.1007/978-1-4612-4964-1_6}}


\bibitem[Qin et~al\mbox{.}(2025)]%
        {qin2025ai}
\bibfield{author}{\bibinfo{person}{Xin Qin}, \bibinfo{person}{Xiang Zhou}, \bibinfo{person}{Chen Chen}, \bibinfo{person}{Dongyuan Wu}, \bibinfo{person}{Hansen Zhou}, \bibinfo{person}{Xiaowei Dong}, \bibinfo{person}{Limei Cao}, {and} \bibinfo{person}{Jackson~G Lu}.} \bibinfo{year}{2025}\natexlab{}.
\newblock \showarticletitle{AI aversion or appreciation? A capability--personalization framework and a meta-analytic review.}
\newblock \bibinfo{journal}{\emph{Psychological bulletin}} \bibinfo{volume}{151}, \bibinfo{number}{5} (\bibinfo{year}{2025}), \bibinfo{pages}{580}.
\newblock


\bibitem[Richter et~al\mbox{.}(2025)]%
        {richter_large_2025}
\bibfield{author}{\bibinfo{person}{Eileen Richter}, \bibinfo{person}{Markus Wolfgang~Hermann Spitzer}, \bibinfo{person}{Annabelle Morgan}, \bibinfo{person}{Luisa Frede}, \bibinfo{person}{Joshua Weidlich}, {and} \bibinfo{person}{Korbinian Moeller}.} \bibinfo{year}{2025}\natexlab{}.
\newblock \showarticletitle{Large language models outperform humans in identifying neuromyths but show sycophantic behavior in applied contexts}.
\newblock \bibinfo{journal}{\emph{Trends in Neuroscience and Education}}  \bibinfo{volume}{39} (\bibinfo{year}{2025}), \bibinfo{pages}{100255}.
\newblock
\showISSN{22119493}
\href{https://doi.org/10.1016/j.tine.2025.100255}{doi:\nolinkurl{10.1016/j.tine.2025.100255}}


\bibitem[Rogers et~al\mbox{.}(2023)]%
        {rogers_too_2023}
\bibfield{author}{\bibinfo{person}{Benjamin~A. Rogers}, \bibinfo{person}{Ovul Sezer}, {and} \bibinfo{person}{Nadav Klein}.} \bibinfo{year}{2023}\natexlab{}.
\newblock \showarticletitle{Too naïve to lead: When leaders fall for flattery.}
\newblock \bibinfo{journal}{\emph{Journal of Personality and Social Psychology}} \bibinfo{volume}{125}, \bibinfo{number}{6} (\bibinfo{year}{2023}), \bibinfo{pages}{1394--1419}.
\newblock
\showISSN{1939-1315, 0022-3514}
\href{https://doi.org/10.1037/pspi0000433}{doi:\nolinkurl{10.1037/pspi0000433}}


\bibitem[Rrv et~al\mbox{.}(2024)]%
        {rrv_chaos_2024}
\bibfield{author}{\bibinfo{person}{Aswin Rrv}, \bibinfo{person}{Nemika Tyagi}, \bibinfo{person}{Md~Nayem Uddin}, \bibinfo{person}{Neeraj Varshney}, {and} \bibinfo{person}{Chitta Baral}.} \bibinfo{year}{2024}\natexlab{}.
\newblock \showarticletitle{Chaos with Keywords: Exposing Large Language Models Sycophancy to Misleading Keywords and Evaluating Defense Strategies}. In \bibinfo{booktitle}{\emph{Findings of the Association for Computational Linguistics: {ACL} 2024}} (Bangkok, Thailand, 2024-08), \bibfield{editor}{\bibinfo{person}{Lun-Wei Ku}, \bibinfo{person}{Andre Martins}, {and} \bibinfo{person}{Vivek Srikumar}} (Eds.). \bibinfo{publisher}{Association for Computational Linguistics}, \bibinfo{pages}{12717--12733}.
\newblock
\href{https://doi.org/10.18653/v1/2024.findings-acl.755}{doi:\nolinkurl{10.18653/v1/2024.findings-acl.755}}


\bibitem[Sharma et~al\mbox{.}(2023)]%
        {sharma_towards_2023}
\bibfield{author}{\bibinfo{person}{Mrinank Sharma}, \bibinfo{person}{Meg Tong}, \bibinfo{person}{Tomasz Korbak}, \bibinfo{person}{David Duvenaud}, \bibinfo{person}{Amanda Askell}, \bibinfo{person}{Samuel~R. Bowman}, \bibinfo{person}{Newton Cheng}, \bibinfo{person}{Esin Durmus}, \bibinfo{person}{Zac Hatfield-Dodds}, \bibinfo{person}{Scott~R. Johnston}, \bibinfo{person}{Shauna Kravec}, \bibinfo{person}{Timothy Maxwell}, \bibinfo{person}{Sam {McCandlish}}, \bibinfo{person}{Kamal Ndousse}, \bibinfo{person}{Oliver Rausch}, \bibinfo{person}{Nicholas Schiefer}, \bibinfo{person}{Da Yan}, \bibinfo{person}{Miranda Zhang}, {and} \bibinfo{person}{Ethan Perez}.} \bibinfo{year}{2023}\natexlab{}.
\newblock \bibinfo{title}{Towards Understanding Sycophancy in Language Models}.
\newblock
\href{https://doi.org/10.48550/ARXIV.2310.13548}{doi:\nolinkurl{10.48550/ARXIV.2310.13548}}


\bibitem[Sherman(2013)]%
        {sherman_selfaffirmation_2013}
\bibfield{author}{\bibinfo{person}{David~K. Sherman}.} \bibinfo{year}{2013}\natexlab{}.
\newblock \showarticletitle{Self‐Affirmation: Understanding the Effects}.
\newblock \bibinfo{journal}{\emph{Social and Personality Psychology Compass}} \bibinfo{volume}{7}, \bibinfo{number}{11} (\bibinfo{year}{2013}), \bibinfo{pages}{834--845}.
\newblock
\showISSN{1751-9004, 1751-9004}
\href{https://doi.org/10.1111/spc3.12072}{doi:\nolinkurl{10.1111/spc3.12072}}


\bibitem[Shin(2020)]%
        {shin_user_2020}
\bibfield{author}{\bibinfo{person}{Donghee Shin}.} \bibinfo{year}{2020}\natexlab{}.
\newblock \showarticletitle{User Perceptions of Algorithmic Decisions in the Personalized {AI} System:Perceptual Evaluation of Fairness, Accountability, Transparency, and Explainability}.
\newblock \bibinfo{journal}{\emph{Journal of Broadcasting \& Electronic Media}} \bibinfo{volume}{64}, \bibinfo{number}{4} (\bibinfo{year}{2020}), \bibinfo{pages}{541--565}.
\newblock
\showISSN{0883-8151, 1550-6878}
\href{https://doi.org/10.1080/08838151.2020.1843357}{doi:\nolinkurl{10.1080/08838151.2020.1843357}}


\bibitem[Skjuve et~al\mbox{.}(2022)]%
        {skjuve_longitudinal_2022}
\bibfield{author}{\bibinfo{person}{Marita Skjuve}, \bibinfo{person}{Asbjørn Følstad}, \bibinfo{person}{Knut~Inge Fostervold}, {and} \bibinfo{person}{Petter~Bae Brandtzaeg}.} \bibinfo{year}{2022}\natexlab{}.
\newblock \showarticletitle{A longitudinal study of human–chatbot relationships}.
\newblock \bibinfo{journal}{\emph{International Journal of Human-Computer Studies}}  \bibinfo{volume}{168} (\bibinfo{year}{2022}), \bibinfo{pages}{102903}.
\newblock
\showISSN{10715819}
\href{https://doi.org/10.1016/j.ijhcs.2022.102903}{doi:\nolinkurl{10.1016/j.ijhcs.2022.102903}}


\bibitem[Sprecher et~al\mbox{.}(2013)]%
        {sprecher2013taking}
\bibfield{author}{\bibinfo{person}{Susan Sprecher}, \bibinfo{person}{Stanislav Treger}, \bibinfo{person}{Joshua~D Wondra}, \bibinfo{person}{Nicole Hilaire}, {and} \bibinfo{person}{Kevin Wallpe}.} \bibinfo{year}{2013}\natexlab{}.
\newblock \showarticletitle{Taking turns: Reciprocal self-disclosure promotes liking in initial interactions}.
\newblock \bibinfo{journal}{\emph{Journal of Experimental Social Psychology}} \bibinfo{volume}{49}, \bibinfo{number}{5} (\bibinfo{year}{2013}), \bibinfo{pages}{860--866}.
\newblock


\bibitem[Sun and Wang(2025)]%
        {sun_be_2025}
\bibfield{author}{\bibinfo{person}{Yuan Sun} {and} \bibinfo{person}{Ting Wang}.} \bibinfo{year}{2025}\natexlab{}.
\newblock \bibinfo{title}{Be Friendly, Not Friends: How {LLM} Sycophancy Shapes User Trust}.
\newblock
\showeprint[arxiv]{2502.10844 [cs]}
\href{https://doi.org/10.48550/arXiv.2502.10844}{doi:\nolinkurl{10.48550/arXiv.2502.10844}}


\bibitem[Sussman et~al\mbox{.}(1980)]%
        {sussman_sex_1980}
\bibfield{author}{\bibinfo{person}{Lyle Sussman}, \bibinfo{person}{Terry~A. Pickett}, \bibinfo{person}{Irene~Anchini Berzinski}, {and} \bibinfo{person}{Frederick~W. Pearce}.} \bibinfo{year}{1980}\natexlab{}.
\newblock \showarticletitle{Sex and sycophancy: Communication strategies for ascendance in same-sex and mixed-sex superior-subordinate dyads}.
\newblock \bibinfo{journal}{\emph{Sex Roles}} \bibinfo{volume}{6}, \bibinfo{number}{1} (\bibinfo{year}{1980}), \bibinfo{pages}{113--127}.
\newblock
\showISSN{0360-0025, 1573-2762}
\href{https://doi.org/10.1007/BF00288366}{doi:\nolinkurl{10.1007/BF00288366}}


\bibitem[Todorov et~al\mbox{.}(2002)]%
        {todorov2002heuristic}
\bibfield{author}{\bibinfo{person}{Alexander Todorov}, \bibinfo{person}{Shelly Chaiken}, {and} \bibinfo{person}{Marlone~D. Henderson}.} \bibinfo{year}{2002}\natexlab{}.
\newblock \showarticletitle{The Heuristic-Systematic Model of Social Information Processing}.
\newblock In \bibinfo{booktitle}{\emph{The Persuasion Handbook: Developments in Theory and Practice}}, \bibfield{editor}{\bibinfo{person}{James~Price Dillard} {and} \bibinfo{person}{Michael Pfau}} (Eds.). \bibinfo{publisher}{Sage}, \bibinfo{address}{Thousand Oaks, CA}, \bibinfo{pages}{195--211}.
\newblock


\bibitem[Trumbo(1999)]%
        {trumbo_heuristicsystematic_1999}
\bibfield{author}{\bibinfo{person}{Craig~W. Trumbo}.} \bibinfo{year}{1999}\natexlab{}.
\newblock \showarticletitle{Heuristic‐Systematic Information Processing and Risk Judgment}.
\newblock \bibinfo{journal}{\emph{Risk Analysis}} \bibinfo{volume}{19}, \bibinfo{number}{3} (\bibinfo{year}{1999}), \bibinfo{pages}{391--400}.
\newblock
\showISSN{0272-4332, 1539-6924}
\href{https://doi.org/10.1111/j.1539-6924.1999.tb00415.x}{doi:\nolinkurl{10.1111/j.1539-6924.1999.tb00415.x}}


\bibitem[Turkle(2011)]%
        {turkle2011tethered}
\bibfield{author}{\bibinfo{person}{Sherry Turkle}.} \bibinfo{year}{2011}\natexlab{}.
\newblock \showarticletitle{The tethered self: Technology reinvents intimacy and solitude.}
\newblock \bibinfo{journal}{\emph{Continuing higher education review}}  \bibinfo{volume}{75} (\bibinfo{year}{2011}), \bibinfo{pages}{28--31}.
\newblock


\bibitem[Usman et~al\mbox{.}(2024)]%
        {usman_persuasive_2024}
\bibfield{author}{\bibinfo{person}{Umair Usman}, \bibinfo{person}{{TaeWoo} Kim}, \bibinfo{person}{Aaron Garvey}, {and} \bibinfo{person}{Adam Duhachek}.} \bibinfo{year}{2024}\natexlab{}.
\newblock \showarticletitle{The Persuasive Power of {AI} Ingratiation: A Persuasion Knowledge Theory Perspective}.
\newblock \bibinfo{journal}{\emph{Journal of the Association for Consumer Research}} \bibinfo{volume}{9}, \bibinfo{number}{3} (\bibinfo{year}{2024}), \bibinfo{pages}{319--331}.
\newblock
\showISSN{2378-1815, 2378-1823}
\href{https://doi.org/10.1086/730280}{doi:\nolinkurl{10.1086/730280}}


\bibitem[Velez and Hanus(2016)]%
        {velez_self-affirmation_2016}
\bibfield{author}{\bibinfo{person}{John~A. Velez} {and} \bibinfo{person}{Michael~D. Hanus}.} \bibinfo{year}{2016}\natexlab{}.
\newblock \showarticletitle{Self-Affirmation Theory and Performance Feedback: When Scoring High Makes You Feel Low}.
\newblock \bibinfo{journal}{\emph{Cyberpsychology, Behavior, and Social Networking}} \bibinfo{volume}{19}, \bibinfo{number}{12} (\bibinfo{year}{2016}), \bibinfo{pages}{721--726}.
\newblock
\showISSN{2152-2715, 2152-2723}
\href{https://doi.org/10.1089/cyber.2016.0144}{doi:\nolinkurl{10.1089/cyber.2016.0144}}


\bibitem[Vonk(2002)]%
        {vonk_self-serving_2002}
\bibfield{author}{\bibinfo{person}{Roos Vonk}.} \bibinfo{year}{2002}\natexlab{}.
\newblock \showarticletitle{Self-serving interpretations of flattery: Why ingratiation works.}
\newblock \bibinfo{journal}{\emph{Journal of Personality and Social Psychology}} \bibinfo{volume}{82}, \bibinfo{number}{4} (\bibinfo{year}{2002}), \bibinfo{pages}{515--526}.
\newblock
\showISSN{1939-1315, 0022-3514}
\href{https://doi.org/10.1037/0022-3514.82.4.515}{doi:\nolinkurl{10.1037/0022-3514.82.4.515}}


\bibitem[Wang and Kantarcioglu(2025)]%
        {wang_ask_2025}
\bibfield{author}{\bibinfo{person}{Chengen Wang} {and} \bibinfo{person}{Murat Kantarcioglu}.} \bibinfo{year}{2025}\natexlab{}.
\newblock \bibinfo{title}{Ask {ChatGPT}: Caveats and Mitigations for Individual Users of {AI} Chatbots}.
\newblock
\showeprint[arxiv]{2508.10272 [cs]}
\href{https://doi.org/10.48550/arXiv.2508.10272}{doi:\nolinkurl{10.48550/arXiv.2508.10272}}


\bibitem[Wang et~al\mbox{.}(2025a)]%
        {wang_when_2025}
\bibfield{author}{\bibinfo{person}{Keyu Wang}, \bibinfo{person}{Jin Li}, \bibinfo{person}{Shu Yang}, \bibinfo{person}{Zhuoran Zhang}, {and} \bibinfo{person}{Di Wang}.} \bibinfo{year}{2025}\natexlab{a}.
\newblock \bibinfo{title}{When Truth Is Overridden: Uncovering the Internal Origins of Sycophancy in Large Language Models}.
\newblock
\href{https://doi.org/10.48550/ARXIV.2508.02087}{doi:\nolinkurl{10.48550/ARXIV.2508.02087}}
\newblock
\shownote{Version Number: 2}.


\bibitem[Wang et~al\mbox{.}(2025b)]%
        {wang_embody_2025}
\bibfield{author}{\bibinfo{person}{Kyra Wang}, \bibinfo{person}{Boon-Kiat Quek}, \bibinfo{person}{Jessica Goh}, {and} \bibinfo{person}{Dorien Herremans}.} \bibinfo{year}{2025}\natexlab{b}.
\newblock \bibinfo{title}{To Embody or Not: The Effect Of Embodiment On User Perception Of {LLM}-based Conversational Agents}.
\newblock
\href{https://doi.org/10.48550/ARXIV.2506.02514}{doi:\nolinkurl{10.48550/ARXIV.2506.02514}}


\bibitem[Weng et~al\mbox{.}(2024)]%
        {weng_controllm_2024}
\bibfield{author}{\bibinfo{person}{Yixuan Weng}, \bibinfo{person}{Shizhu He}, \bibinfo{person}{Kang Liu}, \bibinfo{person}{Shengping Liu}, {and} \bibinfo{person}{Jun Zhao}.} \bibinfo{year}{2024}\natexlab{}.
\newblock \bibinfo{title}{{ControlLM}: Crafting Diverse Personalities for Language Models}.
\newblock
\showeprint[arxiv]{2402.10151 [cs]}
\href{https://doi.org/10.48550/arXiv.2402.10151}{doi:\nolinkurl{10.48550/arXiv.2402.10151}}


\bibitem[Xie and Pentina(2022)]%
        {xie_attachment_2022}
\bibfield{author}{\bibinfo{person}{Tianling Xie} {and} \bibinfo{person}{Iryna Pentina}.} \bibinfo{year}{2022}\natexlab{}.
\newblock \showarticletitle{Attachment Theory as a Framework to Understand Relationships with Social Chatbots: A Case Study of Replika}.
\newblock
\href{https://doi.org/10.24251/HICSS.2022.258}{doi:\nolinkurl{10.24251/HICSS.2022.258}}


\bibitem[Zhang et~al\mbox{.}(2025)]%
        {zhang_dark_2025}
\bibfield{author}{\bibinfo{person}{Renwen Zhang}, \bibinfo{person}{Han Li}, \bibinfo{person}{Han Meng}, \bibinfo{person}{Jinyuan Zhan}, \bibinfo{person}{Hongyuan Gan}, {and} \bibinfo{person}{Yi-Chieh Lee}.} \bibinfo{year}{2025}\natexlab{}.
\newblock \showarticletitle{The Dark Side of {AI} Companionship: A Taxonomy of Harmful Algorithmic Behaviors in Human-{AI} Relationships}. In \bibinfo{booktitle}{\emph{Proceedings of the 2025 {CHI} Conference on Human Factors in Computing Systems}} (Yokohama Japan, 2025-04-26). \bibinfo{publisher}{{ACM}}, \bibinfo{pages}{1--17}.
\newblock
\showISBNx{979-8-4007-1394-1}
\href{https://doi.org/10.1145/3706598.3713429}{doi:\nolinkurl{10.1145/3706598.3713429}}


\bibitem[Zimmerman et~al\mbox{.}(2024)]%
        {zimmerman_humanai_2024}
\bibfield{author}{\bibinfo{person}{Anne Zimmerman}, \bibinfo{person}{Joel Janhonen}, {and} \bibinfo{person}{Emily Beer}.} \bibinfo{year}{2024}\natexlab{}.
\newblock \showarticletitle{Human/{AI} relationships: challenges, downsides, and impacts on human/human relationships}.
\newblock \bibinfo{journal}{\emph{{AI} and Ethics}} \bibinfo{volume}{4}, \bibinfo{number}{4} (\bibinfo{year}{2024}), \bibinfo{pages}{1555--1567}.
\newblock
\showISSN{2730-5953, 2730-5961}
\href{https://doi.org/10.1007/s43681-023-00348-8}{doi:\nolinkurl{10.1007/s43681-023-00348-8}}


\end{thebibliography}
\end{document}